\documentclass[aps,prx,twocolumn,showpacs,psfig,superscriptaddress,longbibliography]{revtex4-1}

\usepackage{times}
\usepackage{graphicx}
\usepackage{float}
\usepackage{latexsym,amsmath,amssymb,bm,euscript}
\usepackage{color}
\usepackage{subfigure}
\usepackage{epstopdf}
\usepackage[colorlinks=true,linkcolor=blue,citecolor=blue]{hyperref}
\usepackage{soul}
\usepackage[normalem]{ulem}
\usepackage{mathrsfs}
\usepackage{amsmath}
\usepackage{lettrine}
\usepackage{bbding}
\usepackage{xspace}
\usepackage{textcomp}
\usepackage{textcase}
\usepackage{setspace}

\newcommand {\B}{\textcolor {blue}}

\newcommand{\rev}[1]{{\color[rgb]{0,0,0}{#1}}}

\newcommand{\Fig}[1]{Fig.~\ref{#1}}

\def\RuCl{\ensuremath{\alpha}-RuCl\ensuremath{_3}\xspace}

\def\para{\ensuremath{/\kern -0.8em /}\xspace}

\def\beqn{\begin{eqnarray}}
\def\eeqn{\end{eqnarray}}
\def\beq{\begin{equation}}
\def\eeq{\end{equation}}

\newcommand{\Beq}{\begin{eqnarray*} }
\newcommand{\Eeq}{\end{eqnarray*} }
\newcommand{\Bmat}{\left(\begin{matrix}}
\newcommand{\Emat}{\end{matrix}\right)}
\newcommand{\up}{\uparrow}
\newcommand{\dn}{\downarrow}

\graphicspath{{./}}

\begin{document}
\title{Identification of Magnetic Interactions and High-field Quantum Spin Liquid in $\alpha$-RuCl$_3$}

\author{Han Li}
\thanks{These authors contributed equally to this work.}
\affiliation{School of Physics, Beihang University, Beijing 100191, China}

\author{Hao-Kai Zhang}
\thanks{These authors contributed equally to this work.}
\affiliation{School of Physics, Beihang University, Beijing 100191, China}
\affiliation{Institute for Advanced Study, Tsinghua University, 
Beijing 100084, China}

\author{Jiucai Wang}
\thanks{These authors contributed equally to this work.}
\affiliation{Institute for Advanced Study, Tsinghua University, 
Beijing 100084, China}
\affiliation{Department of Physics, Renmin University of China, 
Beijing 100872, China}

\author{Han-Qing Wu}
\thanks{These authors contributed equally to this work.}
\affiliation{Center for Neutron Science and Technology, 
School of Physics, Sun Yat-sen University, Guangzhou, 510275, China}

\author{Yuan Gao}
\affiliation{School of Physics, Beihang University, Beijing 100191, China}

\author{Dai-Wei Qu}
\affiliation{School of Physics, Beihang University, Beijing 100191, China}

\author{Zheng-Xin Liu}
\email{liuzxphys@ruc.edu.cn}
\affiliation{Department of Physics, Renmin University of China, 
Beijing 100872, China}

\author{Shou-Shu Gong}
\email{shoushu.gong@buaa.edu.cn}
\affiliation{School of Physics, Beihang University, Beijing 100191, China}
\affiliation{International Research Institute of Multidisciplinary Science, Beihang University, Beijing 100191, China}

\author{Wei Li}
\email{w.li@buaa.edu.cn}
\affiliation{School of Physics, Beihang University, Beijing 100191, China}
\affiliation{International Research Institute of Multidisciplinary Science, Beihang University, Beijing 100191, China}
\affiliation{\rev{Institute of Theoretical Physics, Chinese Academy of Sciences, Beijing 100190, China}}

\begin{abstract} 
\end{abstract}
\date{\today}
\maketitle

\noindent{\bf{Abstract}}\\
{The frustrated magnet $\alpha$-RuCl$_3$ 
constitutes a fascinating quantum material platform that 
harbors the intriguing Kitaev physics. However, a consensus on its 
intricate spin interactions and field-induced quantum phases has not been 
reached yet. Here we exploit multiple state-of-the-art many-body 
methods and determine the microscopic spin model that  
quantitatively explains major observations in \RuCl, 
including the zigzag order, double-peak specific heat, 
magnetic anisotropy, and the characteristic M-star dynamical 
spin structure, etc. According to our model simulations, the in-plane 
field drives the system into the polarized phase at about 7~T
and a thermal fractionalization occurs at finite 
temperature, reconciling observations in different experiments.
Under out-of-plane fields, the zigzag order is suppressed at 35~T, 
above which, and below a polarization field of 100~T level, 
there emerges a field-induced quantum spin liquid. 
The fractional entropy and algebraic low-temperature specific 
heat unveil the nature of a gapless spin liquid, 
which can be explored in high-field measurements on \RuCl.
\\
}

\noindent{\bf{Introduction}}\\
The
spin-orbit magnet \RuCl, with edge-sharing RuCl$_6$ octahedra 
and a nearly perfect honeycomb plane, has been widely believed to 
be a correlated insulator with the Kitaev interaction~\cite{Plumb2014,
Sears2015,Kubota2015,Sandilands2016,Zhou2016,Sears2020}.
The compound \RuCl, and the Kitaev materials in general, 
have recently raised great research interest in exploring the 
inherent Kitaev physics~\cite{Jackeli2009,Trebst2017arXiv,Winter2017,
Takagi2019,Janssen2019}, which can realize non-Abelian anyon 
with potential applications in topological quantum
computations~\cite{Kitaev2003,Kitaev2006}. Due to additional 
non-Kitaev interactions in the material, \RuCl exhibits a zigzag 
antiferromagnetic (AF) order at sufficiently low temperature 
($T_{\rm c}  \simeq 7$~K)~\cite{Sears2015,Banerjee2017,Do2017},
which can be suppressed by an external in-plane field of 7-8~T
\cite{Sears2017,Zheng2017,Banerjee2018}.
Surprisingly, the thermodynamics and the unusual excitation 
continuum observed in the inelastic neutron scattering (INS) 
measurements suggest the presence of fractional excitations 
and the proximity of $\alpha$-RuCl$_3$ to a quantum spin 
liquid (QSL) phase~\cite{Banerjee2016,Banerjee2017,Do2017}.
Furthermore, experimental probes including
the nuclear magnetic resonance (NMR)
\cite{Zheng2017,Baek2017,Jansa2018},
Raman scattering~\cite{Wulferding2020},
electron spin resonance (ESR)~\cite{Ponomaryov2020},
THz spectroscopy ~\cite{Little2017,Wang2017THz},
and magnetic torque~\cite{Leahy2017,Modic2020}, etc, 
have been employed to address the possible Kitaev physics 
in $\alpha$-RuCl$_3$ from all conceivable angles. 
In particular, the unusual (even half-integer quantized) thermal 
Hall signal was observed in a certain temperature and field window 
\cite{Kasahara2018Unusual,Kasahara2018,Yokoi2020arXivHalf,
Yamashita2020sample}, suggesting the emergent Majorana 
fractional excitations. However, significant open questions 
remain to be addressed: whether the in-plane field in \RuCl 
induces a QSL ground state that supports the spin-liquid signals 
in experiment, and furthermore, is there a QSL phase induced 
by fields along other direction?

To accommodate the QSL states in quantum materials {like \RuCl}, 
realization of magnetic interactions of Kitaev type plays a central role. 
Therefore, the very first step toward the precise answer to above 
questions is to pin down an effective low-energy spin model of \RuCl. 
As a matter of fact, people have proposed a number of spin models 
with various couplings~\cite{Winter2016,Winter2017NC,
Wu2018,Cookmeyer2018,Kim2016,Suzuki2018,Suzuki2019,
Ran2017,Wang2017,Ozel2019,Banerjee2016,HSKim2015},
yet even the signs of the couplings are not easy to determine 
and currently no single model can simultaneously cover the 
major experimental observations~\cite{Laurell2020},
leaving a gap between theoretical understanding 
and experimental observations.
In this work, we exploit multiple accurate many-body approaches
to tackle this problem, including the exponential tensor renormalization 
group (XTRG)~\cite{Chen2018, Lih2019} for thermal states, 
the density matrix renormalization group (DMRG) 
and variational Monte Carlo (VMC) for the ground state, 
and the exact diagonalization (ED) for the spectral properties. 
Through large-scale calculations, we determine an effective
Kitaev-Heisenberg-Gamma-Gamma$'$ ($K$-$J$-$\Gamma$-$\Gamma'$) 
model [cf. Eq.~(\ref{Eq:HamRuCl3}) below] that 
can perfectly reproduce the major experimental features 
in the equilibrium and dynamic measurements.

Specifically, in our $K$-$J$-$\Gamma$-$\Gamma'$ model
the Kitaev interaction $K$ is much greater than other 
non-Kitaev terms and found to play the predominant role 
in the intermediate temperature regime, showing  
that $\alpha$-RuCl$_3$ is indeed in close proximity to a QSL.
As the compound, our model also possesses a low-$T$ 
zigzag order, which is melted at about 7~K. At intermediate
energy scale, a characteristic M star in the dynamical spin 
structure is unambiguously reproduced. Moreover, we find 
that in-plane magnetic field suppresses the zigzag order 
at around 7~T, and drives the system into a
trivial polarized phase. Nevertheless, {even above the 
partially polarized states,} our finite-temperature
calculations suggest that $\alpha$-RuCl$_3$ could have 
a fractional liquid regime with exotic Kitaev paramagnetism,
reconciling previous experimental debates.
We put forward proposals to explore the fractional liquid in 
$\alpha$-RuCl$_3$ via thermodynamic and spin-polarized INS 
measurements. Remarkably, when the magnetic field is applied perpendicular
to the honeycomb plane, we disclose a QSL phase driven by high fields, 
which sheds new light on the search of QSL in Kitaev materials. Further,
we propose experimental probes through magnetization and calorimetry
\cite{Imajo2020highresolution} measurements under 100-T class pulsed 
magnetic fields~\cite{Zhou2020particlehole, Zhou2020private}.\\

\noindent{\bf{Results}}\\
\textbf{Effective spin model and quantum many-body methods.} 
We study the $K$-$J$-$\Gamma$-$\Gamma'$ honeycomb model 
with the interactions constrained within the nearest-neighbor sites, i.e.,
\begin{equation}
\begin{split}
H=& \sum_{\langle i,j\rangle_{\gamma}} [K S_i^{\gamma}S_j^{\gamma} + J\,\textbf{S}_i\cdot \textbf{S}_j 
+\Gamma(S_i^{\alpha}S_j^{\beta}+S_i^{\beta}S_j^{\alpha}) \\
& +\Gamma'(S_i^{\gamma}S_j^{\alpha}+S_i^{\gamma}S_j^{\beta}+S_i^{\alpha}S_j^{\gamma}+S_i^{\beta}S_j^{\gamma})],
\end{split}
\label{Eq:HamRuCl3}
\end{equation}
where $\textbf{S}_i = \{ S^x_i, S^y_i, S^z_i \}$ 
are the pseudo spin-$1/2$ operators at site $i$,
and $\langle i,j\rangle_{\gamma}$ denotes the 
nearest-neighbor pair on the $\gamma$ bond, 
with $\{ \alpha, \beta, \gamma\}$ being $\{ x,y,z\}$ 
under a cyclic permutation. $K$ is the Kitaev coupling,
$\Gamma$ and $\Gamma'$ the off-diagonal couplings,
and $J$ is the Heisenberg term.
The symmetry of the model, besides the lattice 
translation symmetries, is described by the finite 
magnetic point group $D_{\rm 3d} \times Z_2^T$, where $Z_2^T = \{E,T\}$ 
is the time-reversal symmetry group and each element in $D_{\rm 3d}$ 
stands for a combination of lattice rotation and spin rotation due to 
the spin-orbit coupling. The symmetry group restricts the physical 
properties of the system. For instance, the Land\'e $g$ tensor and 
the magnetic susceptibility tensor, should be uni-axial.

\begin{figure}[t]
\includegraphics[width=9cm]{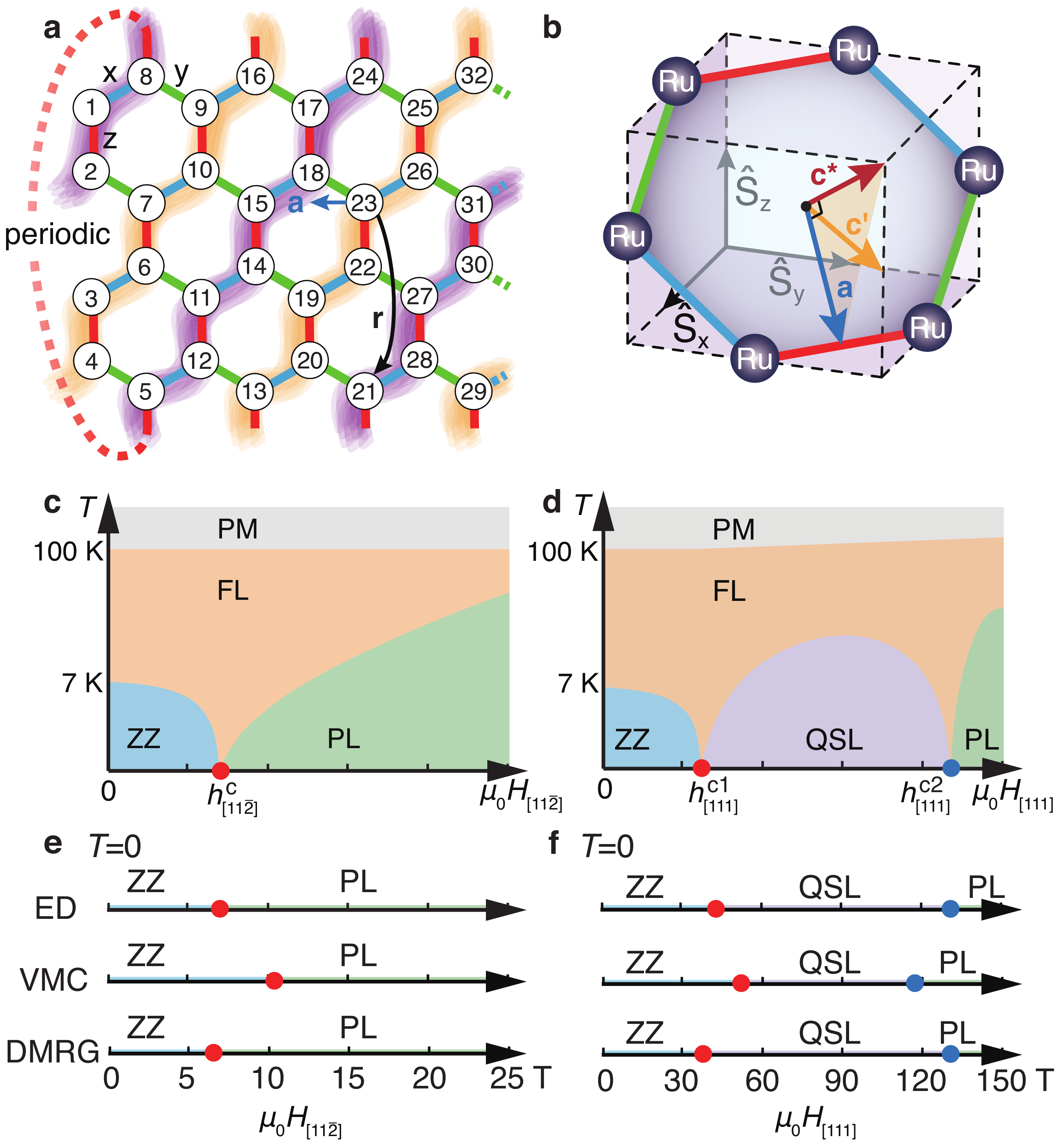}
\renewcommand{\figurename}{\textbf{Fig.}}
\caption{\textbf{Honeycomb lattice, crystalline directions, 
and phase diagrams.} \textbf{a} shows the YC $W \times L \times 2$ 
honeycomb lattice of width $W=4$ and length $L$, 
with two sites per unit cell. The quasi-1D mapping 
path and site ordering is shown by the numbers, 
{with the edge of each small hexagon set as 
length unit.} In the zigzag magnetic order, 
spins align parallel along the highlighted lines, 
while anti-parallel between the yellow and purple lines. 
\textbf{b} shows an unit cell in $\alpha$-RuCl$_3$, 
with the $\textbf{a}$, $\textbf{c}^*$, and $\textbf{c}'$ 
axes indicated. Finite-temperature phase diagrams 
under in-plane ($H_{[11\bar{2}]} \parallel \textbf{a}$) 
and out-of-plane ($H_{[111]} \parallel \textbf{c}^*$) 
fields are presented in \textbf{c} and \textbf{d}, respectively, with in-plane critical field 
$h^c_{[11\bar{2}]} \simeq7$~T (computed at 1.9~K),
and two out-of-plane critical fields $h^{c1}_{[111]}\approx35$~T 
(lower) and $h^{c2}_{[111]} \approx 120$-130~T 
(upper), estimated at 0~K. In the phase diagrams, 
ZZ stands for the zigzag phase, PM for paramagnetic, 
FL for fractional liquid, and PL for the polarized phase, 
with the QPTs (red and blue points) indicated. 
\textbf{e},\textbf{f} show the $T=0$ phase diagrams obtained by
ED, VMC and DMRG, where the in-plane critical field 
is pinpointed at between 6.5 and 10~T, 
while two QPTs and a QSL phase 
are uncovered under out-of-plane fields.
}
\label{Fig:PhsDgr}
\end{figure}

We recall that the $\Gamma'$ term is important for stabilizing 
the zigzag magnetic order at low temperature in the extended 
{ferromagnetic (FM)} Kitaev model {with $K<0$}~\cite{Kim2016,Gordon2019,Lee2020}.
While the zigzag order can also be induced by the 
third-neighbor Heisenberg coupling $J_3$
\cite{Winter2016,Winter2017NC}, we constrain 
ourselves within a minimal $K$-$J$-$\Gamma$-$\Gamma'$ 
model in the present study and leave the discussion 
on the $J_3$ coupling in the Supplementary Note~\B{1}.
In the simulations of \RuCl under magnetic fields, 
we mainly consider the in-plane field along the $[11\bar2]$ direction,
$H_{[11\bar{2}]} \parallel \textbf{a}$, and the out-of-plane field along 
the [111] direction, $H_{[111]} \parallel \textbf{c}^*$, with the 
corresponding Land\'e factors $g_{\rm ab} (= g_{[11\bar{2}]})$ 
and $g_{\rm c^*} (= g_{[111]})$, respectively. 
The index $[l,m,n]$ represents the field direction 
in the spin space depicted in \Fig{Fig:PhsDgr}b.
Therefore, the Zeeman coupling between field $H_{[l,m,n]}$ 
to local moments can be written as 
$H_{\rm Zeeman} =g_{[l,m,n]} \mu_{\rm B} \mu_0 H_{[l,m,n]} S_i^{[l,m,n]}$, 
where $S^{[l,m,n]} \equiv   \mathbf S \cdot \mathbf d_{l,m,n}$ with
$\mathbf S = (S_x, S_y, S_z)$ and $\mathbf d_{l,m,n} = 
(l, m, n)^T/\sqrt{l^2+m^2+n^2}$. The site index $i = 1, \cdots, N$, 
with $N \equiv W \times L \times 2$ the total site number.

In the simulations, various quantum many-body 
calculation methods have been employed (see Methods). 
The thermodynamic properties under zero and finite magnetic 
fields are computed by XTRG on finite-size systems 
(see, e.g, YC4 systems shown in \Fig{Fig:PhsDgr}a).
The model parameters are pinpointed by fitting the XTRG 
results to the thermodynamic measurements, 
and then confirmed by  the ground-state \rev{magnetization} 
calculations by DMRG with the same geometry and VMC 
on an 8$\times$8$\times$2 torus. Moreover, 
the ED calculations of the dynamical properties are performed 
on a 24-site torus, which are in remarkable agreement to 
experiments and further strengthen the validity and accuracy 
of our spin model. Therefore, by combining these {cutting-edge} 
many-body approaches, we explain the experimental 
observations from the determined effective spin Hamiltonian, 
and explore the field-induced QSL in 
$\alpha$-RuCl$_3$ under magnetic fields.

\begin{figure*}[t]
\includegraphics[width=17cm]{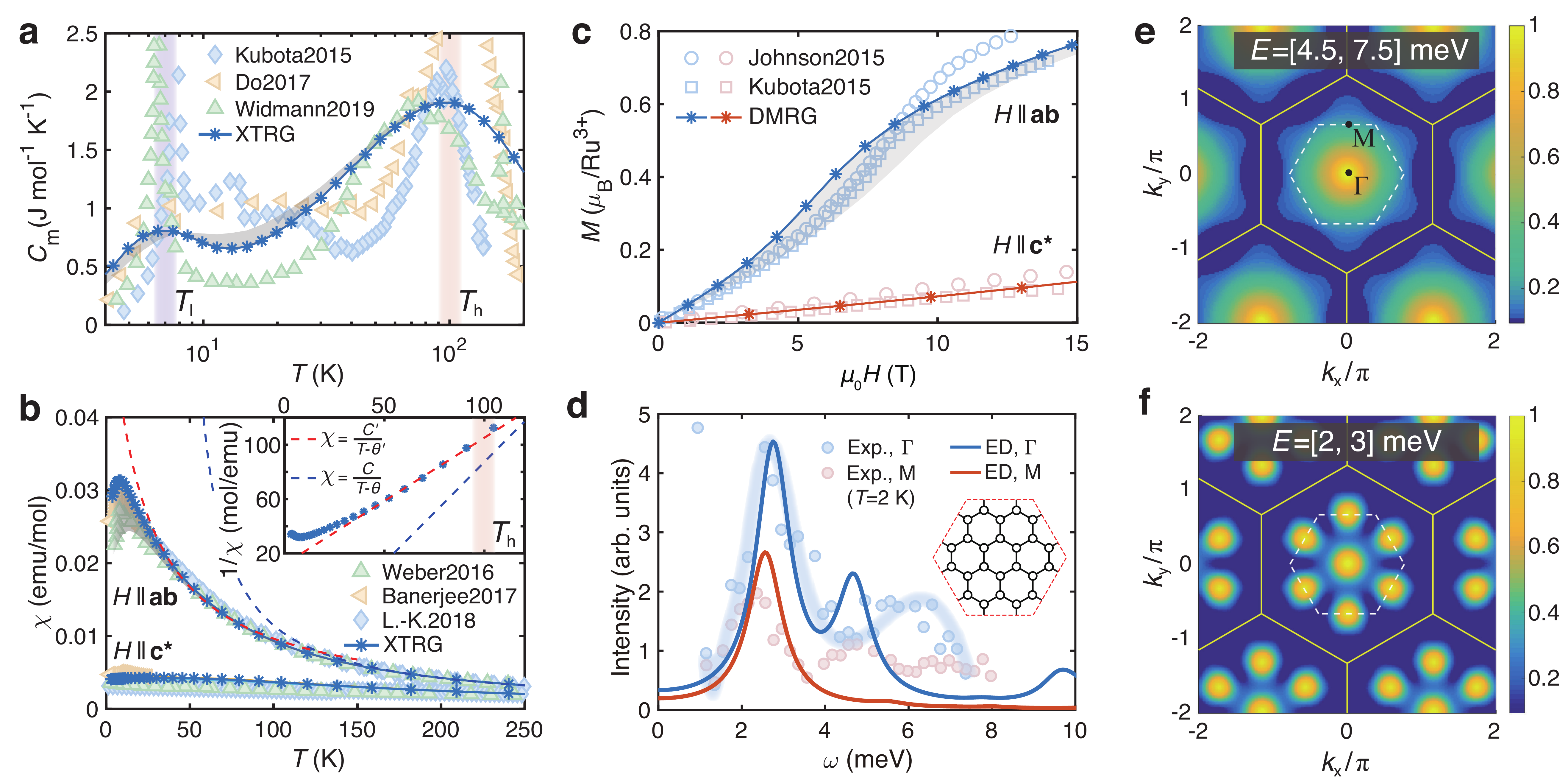}
\renewcommand{\figurename}{\textbf{Fig.}}
\caption{\textbf{Model simulations of thermodynamic 
and dynamic properties.} With the 
$K$-$J$-$\Gamma$-$\Gamma'$ model, we perform many-body 
simulations and compare the results to experiments, 
including \textbf{a} the magnetic specific heat $C_{\rm m}$ 
\cite{Kubota2015,Do2017,Widmann2019},
\textbf{b} in-plane ($\chi_{\rm ab}$) and out-of-plane ($\chi_{\rm c^*}$) 
susceptibilities (measured at the field of 1~T)~\cite{Lampen-Kelley2018,
Banerjee2017,Weber2016}, and \textbf{c} magnetization curves $M(H)$ 
\cite{Johnson2015, Kubota2015} on the {YC4 $\times L \times$2 
lattice, with the grey shaded region indicating the influences of 
different system lengths $L=4,6$.} The Curie-Weiss fittings of 
the high- and intermediate-$T$ susceptibility $\chi_{\rm ab}$ 
in \textbf{b} lead to $C \simeq 0.67$~cm$^3$K/mol, 
$\theta \simeq 41.4$~K, and $C'\simeq1.07$~cm$^3$ K/mol, 
$\theta' \simeq -12.7$~K, respectively. We also compute the 
dynamical spin structure by ED and compare the results with 
neutron scattering intensity $\mathcal{I}(\textbf{k}, \omega)$
at $\textbf{k} =\Gamma$ and M in \textbf{d}, {on a $C_3$ symmetric 
24-site cluster shown in the inset (see more information and its 
comparison to another 24-site cluster with lower symmetry in the 
Supplementary Fig.~\B{4}).} The light blue bold line is a guide to 
the eye for INS data at $\Gamma$ point~\cite{Banerjee2018}.
The calculated intensity peak positions $\omega_\Gamma$ 
and $\omega_{\rm M}$ are in very good agreement with 
experiments.  We further plot the integrated intensities within 
the energy interval \textbf{e} [4.5, 7.5] meV and \textbf{f} [2, 3] meV, 
where a clear M-star shape is reproduced in \textbf{e} 
at intermediate energy and the bright M points are 
evident in \textbf{f} at low energy, both in excellent agreement 
with the INS experiments \cite{Banerjee2017}.
The dashed white hexagon marks the first BZ, 
and the outer yellow hexagon is the extended BZ.
}
\label{Fig:Fittings}
\end{figure*}

\textbf{Model parameters.} 
As shown in Fig.~\ref{Fig:Fittings}a-b, through simulating the 
experimental measurements, including the magnetic specific heat 
and both in- and out-of-plane susceptibility data
\cite{Kubota2015,Do2017,Widmann2019,Lampen-Kelley2018,
Banerjee2017,Weber2016,Johnson2015}, we accurately 
determine the parameters in the Hamiltonian Eq.~(\ref{Eq:HamRuCl3}), 
which read $K=-25$ meV, $\Gamma=0.3\ |K|$, $\Gamma'=-0.02\ |K|$, 
and $J=-0.1\ |K|$. The in- and out-of-plane Land\'e factors are found to 
be $g_{\rm ab}=2.5$ and $g_{\rm c^*}=2.3$, respectively. We find that 
both the magnetic specific heat $C_{\rm m}$ and the two susceptibilities
(in-plane $\chi_{\rm ab}$ and out-of-plane $\chi_{\rm c^*}$) are quite 
sensitive to the $\Gamma$ term, and the inclusion of $\Gamma'$($J$) 
term can significantly change the low-$T$ $C_{\rm m}$($\chi_{\rm ab}$) 
data. Based on these observations, we accurately pinpoint the 
various couplings. The details of parameter determination, with 
detailed comparisons to the previously proposed candidate models 
can be found in Supplementary Notes~\B{1, 2}. {To check the robustness 
and uniqueness of the parameter fittings,} we have also performed 
an automatic Hamiltonian searching~\cite{Yu2020Learning} with the Bayesian 
optimization {combined large-scale thermodynamics solver XTRG}, 
and find that the above effective parameter set indeed
locates within the optimal regime of the optimization (Supplementary 
Note~\B{1}). In addition, the validity of our \RuCl model is firmly supported 
by directly comparing the model calculations to the measured magnetization 
curves in \Fig{Fig:Fittings}c and INS measurements in \Fig{Fig:Fittings}d-f.

In our $K$-$J$-$\Gamma$-$\Gamma'$ model of \RuCl, 
we see a dominating FM Kitaev interaction 
and a sub-leading positive $\Gamma$ term ($\Gamma > 0$), 
which fulfill the interaction signs proposed from recent 
experiments~\cite{Ran2017,Wu2018,Sears2020} 
and agree with some \textit{ab initio} studies
\cite{Kim2016,Winter2017,Winter2017NC,Suzuki2018,Laurell2020}.
The strong Kitaev interaction seems to play a predominant role 
at intermediate temperature, which leads to the fractional liquid 
regime and therefore naturally explains the observed proximate 
spin liquid behaviors~\cite{Banerjee2016,Banerjee2017,Do2017}.

\begin{figure*}[t!]
\includegraphics[angle=0,width=1\linewidth]{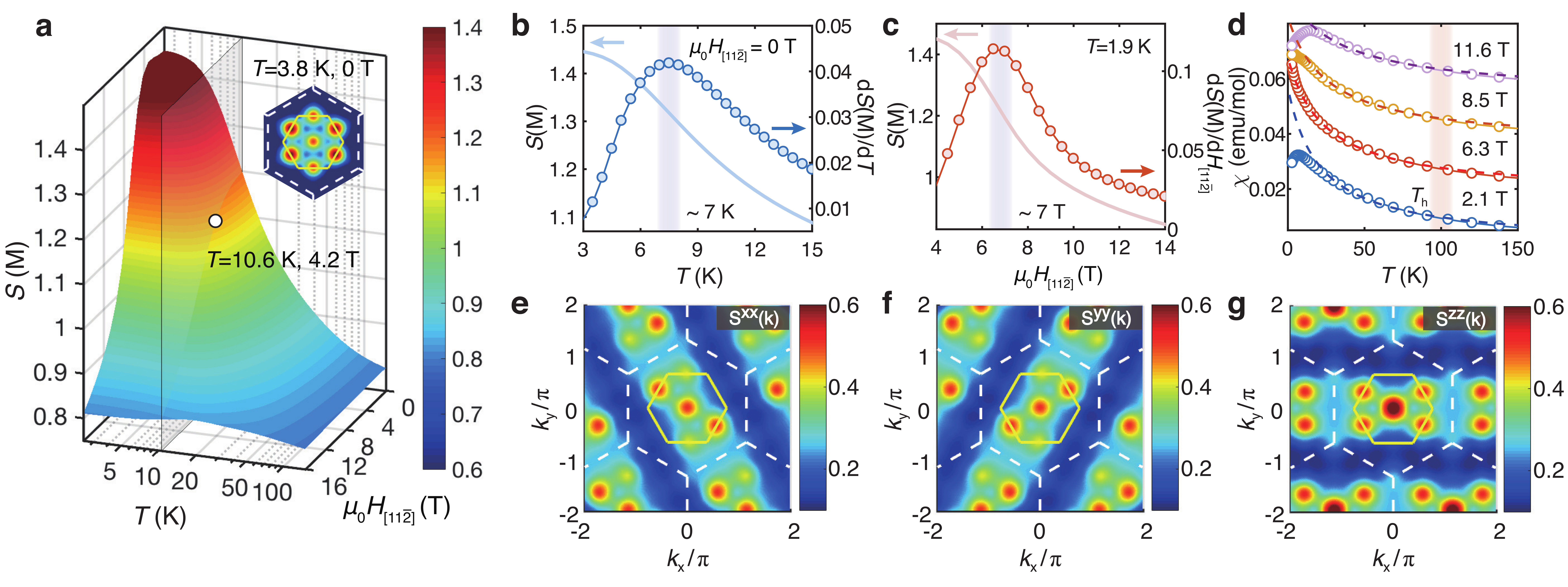}
\renewcommand{\figurename}{\textbf{Fig.}}
\caption{\textbf{Low-temperature zigzag order 
and intermediate-$T$ fractional liquid regime under in-plane 
fields $H_{[11\bar{2}]}\parallel \textbf{a}$.}
\textbf{a} Surface plot of the M-point spin structure 
factor $S(\rm{M})$ (average over six M points in the BZ) 
vs. temperature $T$ and field $H_{[11\bar{2}]}$.
The inset is a low-$T$ structure factor with six bright 
M points in the BZ, representing the zigzag order.
The $S(\rm{M})$ curve and its derivative over $T$ 
and $H_{[11\bar{2}]}$ are shown in \textbf{b} and \textbf{c}, respectively,
which indicate a suppression of zigzag order at around the 
temperature $T_{\rm c} \simeq 7$~K and field $h^c_{[11\bar{2}]} \simeq 7$~T.
\textbf{d} shows the emergent Curie-Weiss behavior in the magnetic 
susceptibility at intermediate $T$, with the fitted $C'\simeq$ 1.17, 1.20, 1.21, 
and 1.27 $\rm{cm^3 K/mol}$ and $\theta' \simeq$ -20.7, -24.2, -26, 
and -34.4~K for fields $\mu_0H_{[11\bar{2}]}=$2.1, 6.3, 8.5 and 11.6~T,
respectively. The different $\chi$ curves under various fields 
are shifted vertically by 0.018 emu/mol for clarify. 
The bright stripes of spin-resolved magnetic structure factors, 
i.e., $S^{xx}(\textbf{k})$, $S^{yy}(\textbf{k})$, and $S^{zz}(\textbf{k})$
as shown in \textbf{e},\textbf{f},\textbf{g}, respectively, 
calculated at $\mu_0H_{[11\bar{2}]}=4.2$~T and $T=10.6$~K, rotate 
counterclockwise (by 120$^\circ$) as the spin component switches, 
reflecting the peculiar bond-oriented spin correlations 
in the intermediate fractional liquid regime.
}
\label{Fig:InPlane}
\end{figure*}

\textbf{Magnetic specific heat and two-temperature scales.} 
We now show our simulations of the $K$-$J$-$\Gamma$-$\Gamma'$
model and compare the results to the thermodynamic measurements.
In \Fig{Fig:Fittings}a, the XTRG results accurately 
capture the prominent double-peak feature of the 
magnetic specific heat $C_{\rm m}$, i.e., 
a round high-$T$ peak at $T_{\rm h}\simeq100$~K 
and a low-$T$ one at $T_{\rm l}\simeq7$ K.
As $T_{\rm h}\simeq100$~K is a relatively high-temperature scale 
where the phonon background needs to be carefully deal with~\cite{Widmann2019}, 
and there exists quantitative difference among the various $C_{\rm m}$ 
measurements in the high-$T$ regime~\cite{Kubota2015,Do2017,Widmann2019}.
Nevertheless, the high-$T$ scale $T_{\rm h}$ itself is relatively stable, 
and in Fig.~\ref{Fig:Fittings}a our XTRG result indeed exhibits 
a high-$T$ peak centered at around 100~K, in good agreement 
with various experiments. Note that the high-temperature crossover 
at $T_{\rm h}$ corresponds to the establishment of short-range 
spin correlations, which can be ascribed to the \rev{emergence of} 
itinerant Majorana fermions~\cite{Nasu2015,Do2017} 
in the fractional liquid picture that we will discuss.
Such a crossover can also be observed in the susceptibilities, which 
deviate the high-$T$ Curie-Weiss law and exhibit an 
intermediate-$T$ Curie-Weiss scaling below $T_{\rm h}$~\cite{Han2020},
as shown in Fig.~\ref{Fig:Fittings}b for $\chi_{\rm ab}$ (the same for $\chi_{\rm c^*}$).

At the temperature $T_{\rm l}\simeq 7$~K, 
the experimental $C_{\rm m}$ curves of \RuCl exhibit a very sharp peak, 
corresponding to the establishment of a zigzag magnetic order
\cite{Kubota2015,Do2017,Widmann2019,Sears2015,Banerjee2017}.
Such a low-$T$ scale can be accurately reproduced by our 
model calculations, as shown in Fig.~\ref{Fig:Fittings}a. 
As our calculations are performed on the cylinders of a finite width,
the height of the $T_{\rm l}$ peak is less prominent than experiments, 
where the transition in the compound \RuCl may be enhanced by 
the inter-layer couplings. Importantly, the location of $T_{\rm l}$ 
fits excellently to the experimental results. Below $T_{\rm l}$ 
our model indeed shows significantly enhanced zigzag 
spin correlation, which is evidenced by the low-energy 
dynamical spin structure in Fig.~\ref{Fig:Fittings}f 
and the low-$T$ static structure in the inset of Fig.~\ref{Fig:InPlane}a.

\textbf{Anisotropic susceptibility and magnetization curves.}
It has been noticed from early experimental studies of \RuCl that 
there exists a very strong magnetic anisotropy in the compound
\cite{Sears2015,Kubota2015,Johnson2015,Weber2016,
Banerjee2017,Lampen-Kelley2018,Sears2020},
which was firstly ascribed to anisotropic Land\'e $g$ factor
\cite{Kubota2015,Johnson2015}, and recently to the existence 
of the off-diagonal $\Gamma$ interaction~\cite{Lampen-Kelley2018,Sears2020}.
We compute the magnetic susceptibilities along two prominent 
field directions, i.e., $H_{[11\bar{2}]}$ and $H_{[111]}$, 
and compare them to experiments \rev{in Fig.~\ref{Fig:Fittings}b}
\cite{Weber2016,Banerjee2017,Lampen-Kelley2018}.
The discussions on different in-plane {and tilted fields} 
are left in the Supplementary Note~\B{3}.

In Fig.~\ref{Fig:Fittings}b, we show that both the in- and 
out-of-plane magnetic  susceptibilities $\chi_{\rm ab}$ 
and $\chi_{\rm c^*}$ can be well fitted 
using our $K$-$J$-$\Gamma$-$\Gamma'$ model, 
with dominant Kitaev $K$, considerable off-diagonal $\Gamma$,
as well as similar in-plane ($g_{\rm ab}$) and out-of-plane ($g_{\rm c^*}$) 
Land\'e factors. Therefore, our many-body simulation results 
indicate that the anisotropic susceptibilities mainly 
originate from the off-diagonal $\Gamma$ coupling 
(cf. Supplementary Fig.~\B{2}), in consistent with the 
resonant elastic X-ray scattering~\cite{Sears2020} 
and susceptibility measurements~\cite{Lampen-Kelley2018}.
Moreover, with the parameter set of $K$, $\Gamma$, $\Gamma'$, 
$J$, $g_{\rm ab}$, and $g_{\rm c^*}$ determined from our thermodynamics 
simulations, we compute the magnetization curves 
$M(H_{[l,m,n]}) = 1/N \sum_{i=1}^N g_{[l,m,n]} \mu_{\rm B} \langle 
S_i^{[l,m,n]} \rangle$ along the $[11\bar{2}]$ and [111] 
directions using DMRG, as shown in \Fig{Fig:Fittings}c. 
The two simulated curves, showing clear magnetic anisotropy, 
are in quantitative agreement with the experimental 
measurements at very low temperature~\cite{Johnson2015,Kubota2015}.

\textbf{Dynamical spin structure and the M star.} 
The INS measurements on \RuCl revealed iconic dynamical structure 
features at low and intermediate energies~\cite{Banerjee2017,Do2017}.
With the determined \RuCl model, we compute the 
dynamical spin structure factors using ED, 
and compare the results to experiments.
Firstly, we show in Fig.~\ref{Fig:Fittings}d the constant 
$\textbf{k}$-cut at the $\textbf{k}=\Gamma$ 
and M points (as indicated in Fig.~\ref{Fig:Fittings}e), 
where a quantitative agreement between theory 
and experiment can be observed. In particular, 
the positions of the intensity peak 
$\omega_\Gamma = 2.69\pm0.11$~meV 
and $\omega_{\rm M}=2.2\pm 0.2$ meV 
from the INS measurements~\cite{Banerjee2017},
are accurately reproduced with our determined model. 
For the $\Gamma$-point intensity, the double-peak structure,
which was \rev{observed in experimental measurements}~\cite{Banerjee2018},
can also be well captured.

We then integrate the INS intensity $\mathcal{I}(\textbf{k},
\omega)$ over the low- and intermediate-energy regime 
with the atomic form factor taken into account,
and check their $\textbf{k}$-dependence in \Fig{Fig:Fittings}e-f. 
In experiment, a structure factor with bright $\Gamma$ and 
M points was observed at low energy, and, on the other hand, 
a renowned six-pointed star shape (dubbed M star~\cite{Laurell2020}) 
was reported at intermediate energies~\cite{Banerjee2017,Do2017}.
In Fig.~\ref{Fig:Fittings}e-f, these two characteristic 
dynamical spin structures are reproduced, in exactly 
the same energy interval as experiments. Specifically, 
the zigzag order at low temperature is reflected in the bright 
M points in the Brillouin zone (BZ) when integrated over [2, 3] meV, 
and the $\Gamma$ point in the BZ is also turned on. 
As the energy interval increases to [4.5, 7.5] meV, 
the M star emerges as the zigzag correlation is weakened 
while the continuous dispersion near the $\Gamma$ point
remains prominent. The round $\Gamma$ peak, 
which also appears in the pure Kitaev model, 
is consistent with the strong Kitaev term in our \RuCl model.

\begin{figure}[t!]
\includegraphics[angle=0,width=1\linewidth]{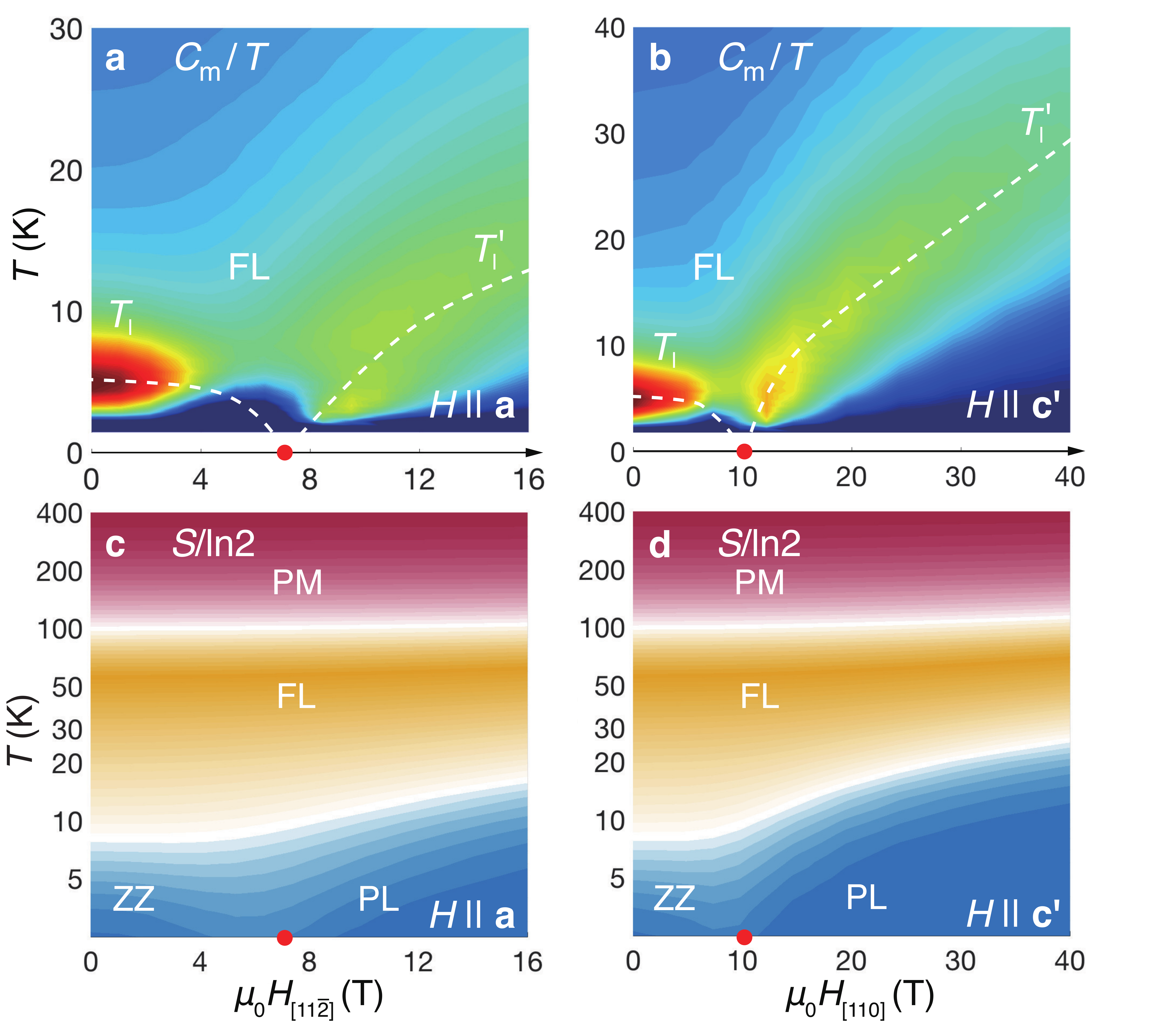}
\renewcommand{\figurename}{\textbf{Fig.}}
\caption{\textbf{Contour plots of $C_{\rm m}/T$ 
and the isentropes.} The magnetic specific heat $C_{\rm m} / T$ 
under $H_{[11\bar{2}]} \parallel \textbf{a}$ and $H_{[110]} \parallel \textbf{c}'$ 
are shown in panels \textbf{a} and \textbf{b}, respectively. The white dashed lines 
indicate the low-temperature scale $T_{\rm l}$ and $T_{\rm l}'$ 
and are guides for the eye. The red dots on the $T=0$ axis
denote the QPT at $h_{[11\bar{2}]}^c \simeq 7$ T 
and $h_{[110]}^c \simeq 10$ T. The isentropes along 
the two field directions are shown in panels \textbf{c} and \textbf{d},
where we find an intermediate regime with fractional 
thermal entropies $\sim (\ln{2}) / 2$.
}
\label{Fig:CmoverT}
\end{figure}

\textbf{Suppressing the zigzag order by in-plane fields.}
In experiments, the low-$T$ zigzag magnetic order
has been observed to be suppressed by the in-plane magnetic 
fields above 7-8~T~\cite{Kubota2015,Johnson2015,
Zheng2017,Banerjee2018}. We hereby investigate this
field-induced effect by computing the spin structure
factors under finite fields. The M-point peak of the structure 
factor $S(\rm M)$ in the $T$-$H_{[11\bar{2}]}$ plane 
characterizes the zigzag magnetic order as shown in 
\Fig{Fig:InPlane}a. The derivatives {$\frac{dS({\rm M})}{dT}|_{H=0}$ 
and $\frac{dS({\rm M})}{dH}|_{T=1.9~\rm{K}}$} are calculated 
in \Fig{Fig:InPlane}b-c, which can only show a round peak 
at the transition as limited by our finite-size simulation.
For $H=0$, the turning temperature is at about $7$~K,  
below which the zigzag order builds up; on the other hand, 
the isothermal $M$-$H$ curves in \Fig{Fig:InPlane}c 
suggest a transition point at $h_{[11\bar{2}]}^c = \mu_0H_{[11\bar{2}]} 
\simeq 7$ T, beyond which the zigzag order is suppressed. 
Correspondingly, in Fig.~\ref{Fig:CmoverT}c [and also in Fig.~\ref{Fig:CmoverT}d],
the low-temperature scale $T_{\rm l}$ decreases as the fields increase,
initially very slow for small fields and then quickly approaches
zero only in the field regime near the critical point, again in very good 
consistency with experimental measurements~\cite{Kubota2015,Zheng2017,Baek2017}.

Besides, from the contour plots of $C_{\rm m}/T$ and the isentropes
in Fig.~\ref{Fig:CmoverT}a, c, one can also recognize the critical
temperature and field consistent with the above estimations. 
Moreover, when the field direction is tilted about $55^{\circ}$ 
away from the $\textbf{a}$ axis in the $\textbf{a}$-$\textbf{c}^*$ 
plane (i.e., $H_{[110]}$ along the $c'$ axis), as shown in
Fig.~\ref{Fig:CmoverT}b, d, our model calculations suggest 
a critical field $h_{[110]}^c = \mu_0 H_{[110]} \simeq 10$~T 
with suppressed zigzag order, in accordance with recent NMR
probe~\cite{Baek2017}. Overall, the excellent agreements of the 
finite-field simulations with different experiments further 
confirm our $K$-$J$-$\Gamma$-$\Gamma'$ model as an 
accurate description of the Kitaev material \RuCl.

\textbf{Finite-$T$ phase diagram under in-plane fields.}
Despite intensive experimental and theoretical studies, 
the phase diagram of \RuCl under in-plane fields remains 
an interesting open question. The thermal Hall
\cite{Kasahara2018}, Raman scattering~\cite{Wulferding2020},
and thermal expansion~\cite{Gass2020} measurements
suggest the existence of an intermediate QSL phase
between the zigzag and polarized phases.
On the other hand, the magnetization~\cite{Kubota2015,Johnson2015}, 
INS~\cite{Banerjee2018}, NMR~\cite{Zheng2017,Baek2017,Jansa2018}, 
ESR~\cite{Ponomaryov2020}, Gr\"uneisen parameter~\cite{Bachus2020GP}, 
and magnetic {torque} measurements~\cite{Modic2020} support 
a single-transition scenario (leaving aside the transition 
between two zigzag phases due to different inter-layer stackings~\cite{Balz2019Neutron}).
{Nevertheless, most experiments found signatures 
of fractional excitations at finite temperature, although 
an alternative multi-magnon interpretation has also 
been proposed~\cite{Ponomaryov2020}.}

\begin{figure*}[t]
\includegraphics[angle=0,width=1\linewidth]{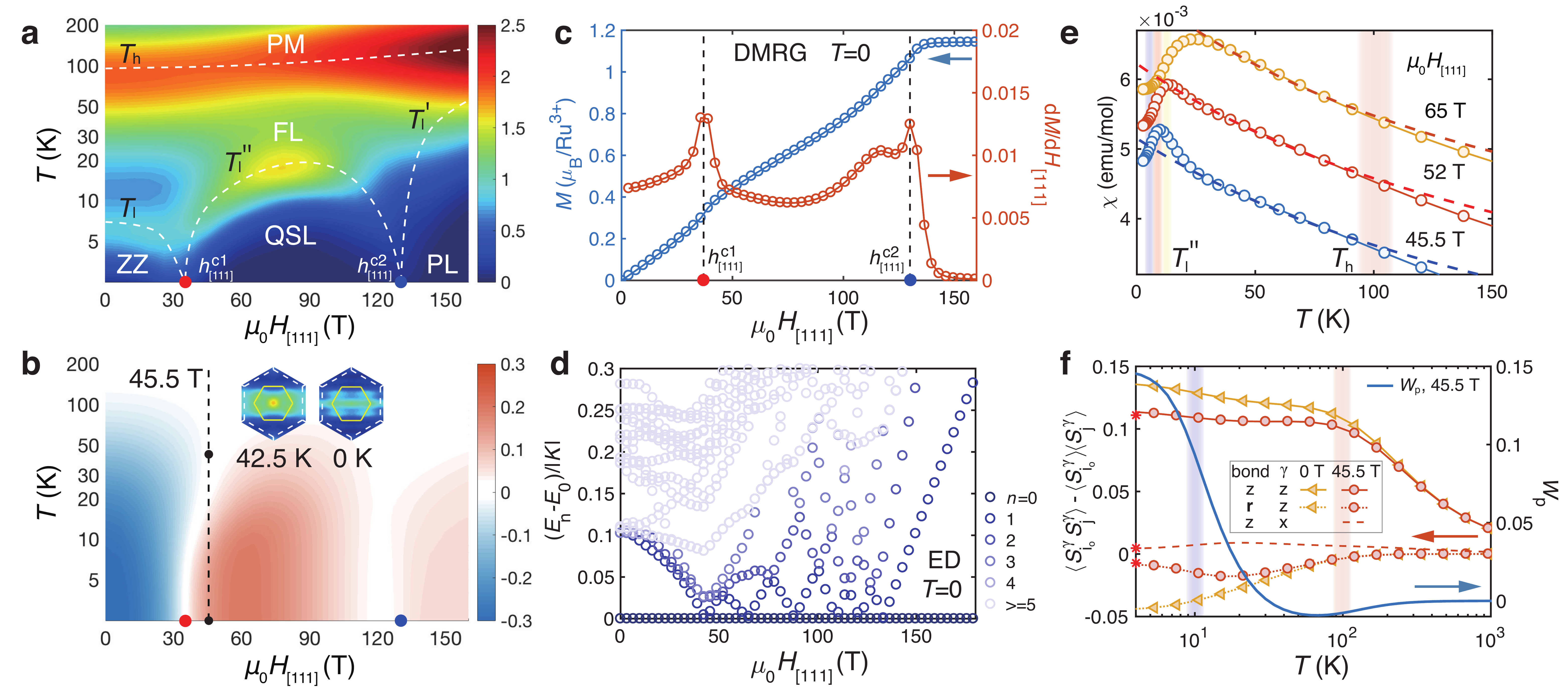}
\renewcommand{\figurename}{\textbf{Fig.}}
\caption{\textbf{Quantum spin liquid regime
of \RuCl model under the out-of-plane
field $H_{[111]} \parallel$ \textbf{c}$^*$.} 
\textbf{a} Color map of the specific heat $C_{\rm m}$,
where a double-peak feature appears at both low and
intermediate fields. The white dashed lines, with 
$T_{\rm h}$, $T_{\rm l}$, and $T''_{\rm l}$ determined 
from the specific heat and $T'_{\rm l}$ from spin structure 
(cf. Supplementary Note~\B{4}), represent 
the phase boundaries and are guides for the eye.
\textbf{b} shows the $Z_2$ flux $W_{\rm p}$, 
where the regime with positive(negative) $W_{\rm p}$ 
is plotted in orange(blue). The zero and finite-temperature
spin structure factors $\tilde{S}^{z z}(\textbf{k})$ (see main text)
at $\mu_0H_{[111]}=45.5$~T are shown in the inset of \textbf{b} and
plotted with the same colorbar as that in \Fig{Fig:InPlane}e-g.
\textbf{c} shows the $T=0$ magnetization curve and its 
derivative d$M$/d$H_{[111]}$, where the two peaks 
can be clearly identified at $\mu_0H_{[111]}\simeq$~35 
and 130~T, respectively. The two transition fields, 
marked by $h_{[111]}^{c1}$ and $h_{[111]}^{c2}$, 
are also indicated in panels \textbf{a} and \textbf{b}.
\textbf{d} ED Energy spectra $E_{\rm n} {\rm -} E_{\rm 0}$ 
under $H_{[111]}$ fields, with gapless low-energy 
excitations  emphasized by dark colored symbols.
\textbf{e} The magnetic susceptibility with an intermediate-$T$ 
Curie-Weiss behavior, with the fitted $C'\simeq$ 1.22, 1.14, 
and 1.00 $\rm{cm^3 K/mol}$ and $\theta' \simeq$ -236.8, 
-217.8, and -187.4~K, for fields $\mu_0H_{[111]}=$45.5, 52 
and 65~T, respectively. The susceptibility curves in \textbf{e} are shifted 
vertically by a constant 0.001~emu/mol for clarify. 
\textbf{f} shows the temperature dependence 
of the spin correlations (left $y$-axis) and the flux $W_{\rm p}$ 
(right $y$-axis), where the correlations 
$\langle S_{i_0}^{z} S_j^{z}\rangle$- $\langle 
S^{z}_{i_0}\rangle\langle S^{z}_{j}\rangle$ 
are measured on both $z$- and $\textbf{r}$-bonds 
[see Fig.~\ref{Fig:PhsDgr}a], 
and $\langle S_{i_0}^{x} S_j^{x}\rangle$- $\langle 
S^{x}_{i_0}\rangle\langle S^{x}_{j}\rangle$  
on the $z$-bond, under two different fields of 0 and 45.5~T
{(with $i_0$ a fixed central reference site)}.
The low-$T$ XTRG data are shown to converge to the 
$T=0$ DMRG results, with the latter marked 
as red asterisks on the left vertical axis.
The flux expectation $W_{\rm p}$ are measured under 
$\mu_0H_{[111]}=45.5$~T, which increases rapidly 
around low-$T$ scale $T''_{\rm l}$.
}
\label{Fig:PD111}
\end{figure*}

Now with the accurate \RuCl model and multiple 
many-body computation techniques, we aim to determine
the phase diagram and nature of the field-driven phase(s). 
Our main results are summarized in Fig.~\ref{Fig:PhsDgr}c, e, 
where a single quantum phase transition (QPT) 
is observed as the in-plane fields $H_{[11\bar{2}]}$ 
increases. Both VMC and DMRG calculations find a trivial 
polarized phase in the large-field side ($\mu_0 H_{[11\bar{2}]}>h_{[11\bar{2}]}^c$), 
as evidenced by the magnetization 
curve in \Fig{Fig:Fittings}c as well as the results in 
Supplementary Note~\B{3}. 

Despite the QSL phase is absent under in-plane fields, 
we nevertheless find a Kitaev fractional liquid at finite 
temperature, whose properties are determined by the 
fractional excitations of the system. For the pure Kitaev 
model, it has been established that the itinerant Majorana 
fermions and $Z_2$ fluxes each releases half 
of the entropy at two crossover temperature 
scales \cite{Nasu2015,Do2017}. Such an intriguing 
regime is also found robust in the extended Kitaev 
model with additional non-Kitaev couplings~\cite{Han2020}. 
Now for the realistic \RuCl model in Eq.~(\ref{Eq:HamRuCl3}), 
we find again the presence of fractional liquid at intermediate $T$. 
As shown in Fig.~\ref{Fig:Fittings}b (zero field) and 
Fig.~\ref{Fig:InPlane}d (finite in-plane fields), the intermediate-$T$ 
Curie-Weiss susceptibility can be clearly observed, 
with the fitted Curie constant $C'$ distinct from the high-$T$ 
paramagnetic constant $C$. This indicates the emergence 
of a novel correlated paramagnetism --- Kitaev paramagnetism --- 
in the material \RuCl. The fractional liquid constitutes an exotic 
finite-temperature quantum state with disordered fluxes and 
itinerant Majorana fermions, driven by the strong Kitaev interaction 
that dominates the intermediate-$T$ regime~\cite{Han2020}.

In Fig.~\ref{Fig:CmoverT}c-d of the isentropes, we find that the Kitaev 
fractional liquid regime is rather broad under either in-plane 
($H_{[11\bar{2}]}$) or tilted ($H_{[110]}$) fields. 
When the field is beyond the critical value, 
the fractional liquid regime gradually gets narrowed, 
from high-$T$ scale $T_{\rm h}$ down to a new 
lower temperature scale $T'_{\rm l}$, below which 
the field-induced uniform magnetization builds up
(see Supplementary Note~\B{3}).
From the specific heat and isentropes in Fig.~\ref{Fig:CmoverT}, 
we find in the polarized phase $T'_{\rm l}$ increases linearly 
as field increases, suggesting that such a low-$T$ scale
can be ascribed to the Zeeman energy. At the intermediate 
temperature, the thermal entropy is around $(\ln{2})/2$ 
[see Fig.~\ref{Fig:CmoverT}c-d], 
indicating that ``one-half" of the spin degree of freedom,
mainly associated with the itinerant Majorana fermions, 
has been gradually frozen below $T_{\rm h}$.
 
Besides, we also compute the spin structure factors 
{$S^{\gamma\gamma}(\textbf{k}) = {1}/{N_{\rm Bulk}}
\, \sum_{i,j\in {\rm Bulk}} e^{ i \textbf{k} 
(\textbf{r}_j-\textbf{r}_{i})} \langle S^{\gamma}_{i} 
S^{\gamma}_j \rangle$} under an in-plane field of 
$\mu_0H_{[11\bar{2}]}=4.2$~T in the fractional liquid regime, 
where {${N_{\rm Bulk}}$ is the number of bulk sites 
(with left and right outmost columns skipped), 
$i,j$ run over the bulk sites}, and $\gamma=x,y,z$. 
{Except for the bright spots at $\Gamma$ and M points,
there appears stripy background in \Fig{Fig:InPlane}e-g 
very similar to that observed in the pure Kitaev 
model~\cite{Han2020}, which reflects the extremely 
short-range and bond-directional spin correlations there.} 
The stripe rotates as the spin component $\gamma$ switches,
because the $\gamma$-type spin correlations $\langle S^{\gamma}_i 
S^{\gamma}_j \rangle_\gamma$ are nonzero only on the 
nearest-neighbor $\gamma$-type bond. As indicated in 
the realistic model calculations, we propose such distinct 
features in $S^{\gamma \gamma}(\textbf{k})$ can be observed 
in the material \RuCl via the polarized neutron diffusive scatterings.

\textbf{Signature of Majorana fermions and the Kitaev fractional liquid.}
It has been highly debated that whether there exists a QSL phase
under intermediate in-plane fields. Although more recent experiments 
favor the single-transition scenario~\cite{Bachus2020GP,Modic2020}, 
there is indeed signature of fractional Majorana fermions and spin liquid 
observed in the intermediate-field regime
\cite{Banerjee2018,Kasahara2018,Jansa2018,Wulferding2020,Modic2020}. 
Based on the model simulations, here we show that our finite-$T$ phase 
diagram in Fig.~\ref{Fig:PhsDgr} provides a consistent scenario that 
reconciles these different in-plane field experiments.

For example, large~\cite{Kasahara2018Unusual,Hentrich2019Large} 
or even half-quantized thermal Hall conductivity was observed 
at intermediate fields and between 4-6~K
\cite{Kasahara2018Quantized,Yamashita2020sample,Yokoi2020arXivHalf}. 
However, it has also been reported that the 
thermal Hall conductivity vanishes rapidly when the field 
\rev{further varies} or the temperature lowers 
below \rev{approximately 2~K}~\cite{Ong2021arXiv}.
Therefore, one possible explanation, according to our model calculations, 
is that the ground state under in-plane fields above $7$~T 
is a trivial polarized phase [see \Fig{Fig:PhsDgr}c], 
while the large thermal Hall conductivity at intermediate 
fields may originate from the Majorana fermion excitations 
in the finite-$T$ fractional liquid
\cite{Nasu2015}. 

In the intermediate-$T$ fractional liquid regime, the Kitaev interaction 
is predominating and the system resembles a pure Kitaev model
under external fields and at a finite temperature. This effect is 
particularly prominent as the field approaches the intermediate regime, 
i.e., near the quantum critical point, where the fractional liquid 
can persist to much lower temperature.
Matter of fact, given a fixed low temperature, when the field is too small or too large,
the system leaves the fractional liquid regime (cf. \Fig{Fig:CmoverT}) 
and the signatures of the fractional excitation become blurred, 
as if there were a finite-field window of ``intermediate spin liquid phase".
Such fractional liquid constitutes a Majorana metal state with a Fermi surface
\cite{Nasu2015,Han2020}, accounting possibly for the observed quantum 
oscillation in longitudinal thermal transport~\cite{Ong2021arXiv}.
Besides thermal transport, the fractional liquid dominated by fractional
excitations can lead to rich experimental ramifications, e.g.,
the emergent Curie-Weiss susceptibilities in susceptibility measurements 
[see Fig.~\ref{Fig:InPlane}d] and the stripy spin structure background 
in the spin-resolved neutron or resonating X-ray scatterings
[Fig.~\ref{Fig:InPlane}e-g], which can be employed to 
probe the finite-$T$ fractional liquid in the compound \RuCl. \\

\begin{figure}[h!]
\includegraphics[angle=0,width=1\linewidth]{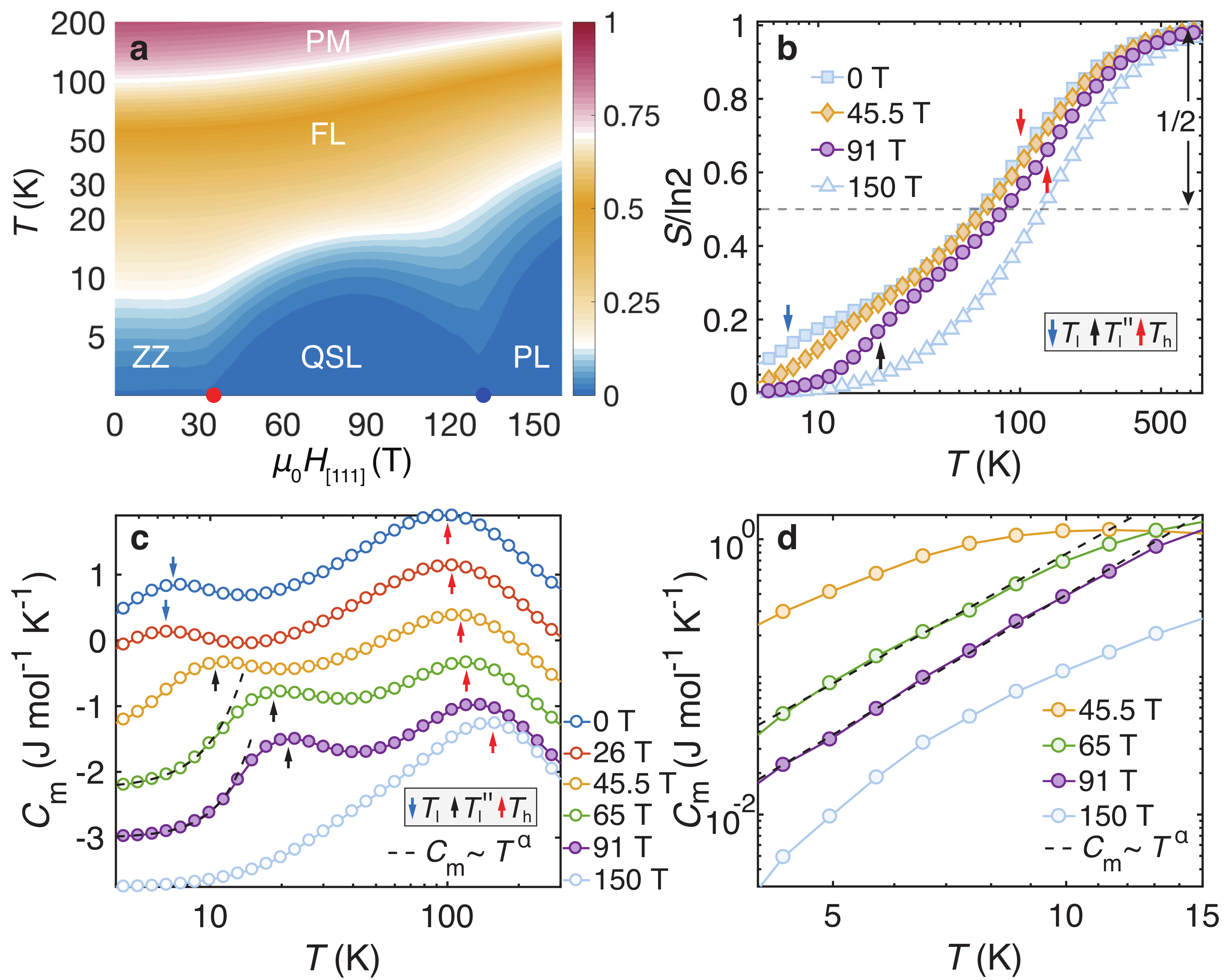}
\renewcommand{\figurename}{\textbf{Fig.}}
\caption{\textbf{Fractional entropy and specific 
heat under out-of-plane fields.}
\textbf{a} The contour plot of thermal entropy $S$/ln2, 
where the two critical fields (red and blue dots) 
are clearly signaled by the two dips in the isentropic 
lines. The entropy curves $S(T)$ are shown in \textbf{b},
where half of the entropy $\Delta S = (\ln{2})/2$ is 
released around the high temperature scale $T_{\rm h}$, 
while the rest half released either by forming zigzag order 
(below $h_{[111]}^{c1}$, see, e.g., the 0~T curve), 
or freezing $Z_2$ flux (between $h_{[111]}^{c1}$ 
and $h_{[111]}^{c2}$, e.g., the 45.5~T and 91~T curves), 
at the lower-temperature scale $T_{\rm l}$ 
and $T''_{\rm l}$ denoted by the arrows. 
For field above $h_{[111]}^{c2}$ the system gradually 
crossover to the polarized states at low temperature.
\textbf{c} shows the specific heat $C_{\rm m}$ curves under
various fields, with the blue(black) arrows indicating 
the low-$T$ scales $T_{\rm l}$($T''_{\rm l}$), and the 
red arrows for $T_{\rm h}$ which increases as the 
field enhances. The $C_{\rm m}$ curves are shifted 
vertically by a constant 0.75~J mol$^{-1}$ K$^{-1}$ for clarify. 
In the QSL regime, the low-$T$ specific heat show 
algebraic behaviors in \textbf{c}, also zoomed in and plotted
with log-log scale in \textbf{d}, and we plot a power-law 
$C_{\rm m} \sim T^{\alpha}$ in both panels as guides 
for the eye, with $\alpha$ estimated to be around 
2.8-3.4 based on our YC4 calculations.
}
\label{Fig:CvS111}
\end{figure}

\textbf{Quantum spin liquid induced by out-of-plane fields.} 
Now we apply the $H_{[111]} \parallel \textbf{c}^*$ field 
out of the plane and investigate the field-induced quantum 
phases in \RuCl. As shown in the phase diagram in 
Fig.~\ref{Fig:PhsDgr}d, f, under the $H_{[111]}$ fields 
a field-induced QSL phase emerges at intermediate fields
between the zigzag and the polarized phases, 
confirmed in both the thermal and ground-state calculations.
The existence of two QPTs and an intermediate phase 
can also be \rev{seen} from the color maps of $C_{\rm m}$, 
$Z_2$ fluxes, and thermal entropies shown 
in Figs.~\ref{Fig:PD111}a-b and \ref{Fig:CvS111}a.

To accurately nail down the two QPTs, we plot the 
$\textbf{c}^*$ DMRG magnetization curve $M(H_{[111]})$
and the derivative $dM/dH_{[111]}$ in \Fig{Fig:PD111}c, 
together with the ED energy spectra in \Fig{Fig:PD111}d, 
from which the ground-state phase diagram can 
be determined (cf. Fig.~\ref{Fig:PhsDgr}f). 
In particular, the lower transition field, 
$h_{[111]}^{c1} \simeq 35$~T, estimated from both 
XTRG and DMRG, is in excellent agreement 
with recent experiment through measuring the 
magnetotropic coefficient~\cite{Modic2020}.
The existence of the upper critical field $h_{[111]}^{c2}$ 
at 100~T level can also be probed with current pulsed 
high field techniques~\cite{Zhou2020private}.

Correspondingly, we find in \Fig{Fig:PD111}b
that the $Z_2$ flux $W_{\rm p}= 
2^6 \langle S_i^x S_j^y S_k^z S_l^x S_m^y 
S_n^z \rangle$ (where $i,j,k,l,m,n$ denote the 
six vertices of a hexagon $p$) changes its sign 
from negative to positive at $h_{[111]}^{c1}$, 
then to virtually zero at $h_{[111]}^{c2}$, and finally 
converge to very small (positive) values in the 
polarized phase. These observations of flux signs in 
different phases are consistent with recent DMRG 
and tensor network studies on a $K$-$\Gamma$-$\Gamma'$
model~\cite{Gordon2019,Lee2020}, {and it is noteworthy 
that the flux is no longer a strictly conserved quantity 
as in the pure Kitaev model, the low-$T$ expectation 
values of $|W_{\rm p}|$ \rev{thus} would be very close to $1$ only 
deep in the Kitaev spin liquid phase
\cite{Gordon2019,Trebst2019,Han2020}.}

From the $C_{\rm m}$ color map in 
Fig.~\ref{Fig:PD111}a and $C_{\rm m}$ curves in 
Fig.~\ref{Fig:CvS111}c, we find double-peaked 
specific heat curves in the QSL phase, 
which clearly indicate the two temperature scales, 
e.g., $T_{\rm h} \simeq 105$~K and $T''_{\rm l} \simeq 10$~K 
for $\mu_0 H_{[111]} = 45.5$~T. They correspond 
to the establishment of spin correlations at $T_{\rm h}$ 
and the alignment of $Z_2$ fluxes at $T''_{\rm l}$, 
respectively, as shown in \Fig{Fig:PD111}f. As a result, 
in Fig.~\ref{Fig:CvS111}a-b the system releases 
$(\ln{2}) / 2$ thermal entropy around $T_{\rm h}$, 
and the rest half is released at around $T''_{\rm l}$.
The magnetic susceptibility curves in Fig.~\ref{Fig:PD111}e 
fall into an intermediate-$T$ Curie-Weiss behavior 
below $T_{\rm h}$ when the spin correlations are established, 
and deviate such emergent universal behavior {when approaching} 
$T''_{\rm l}$ as the gauge degree of freedom 
(flux) {gradually freezes}.

Furthermore, we study the {properties} 
of the QSL phase.
We find very peculiar spin correlations as evidenced 
by the (modified) structure factor 
\begin{equation}
\tilde{S}^{zz}(\textbf{k})=\frac{1}{N_{{\rm Bulk}}}\sum_{i,j\in {\rm Bulk}} 
e^{ i \textbf{k} (\textbf{r}_j-\textbf{r}_{i})} 
(\langle S_{i}^z S_j^z\rangle - \langle 
S^{z}_{i}\rangle\langle S^{z}_{j}\rangle).
\end{equation}
As shown in the inset of \Fig{Fig:PD111}b, 
there appears no prominent peak in $\tilde{S}^{zz}(\textbf{k})$
at both finite and zero temperatures, for a typical 
field of $45.5$~T in the QSL phase. 
{As shown in Fig.~\ref{Fig:PD111}f, 
we find the dominating nearest-neighboring correlations at $T>0$ 
are bond-directional and the longer-range correlations are 
rather weak, same for the QSL ground state at $T=0$.
This is in sharp contrast to the spin correlations in the zigzag 
phase.} More spin structure results computed with both XTRG 
and DMRG can be found in Supplementary Note~\B{4}. 
Below the temperature $T''_{\rm l}$, we observe an algebraic 
specific heat behavior as shown in Fig.~\ref{Fig:CvS111}c-d, 
which strongly suggests a gapless QSL. We remark 
that the gapless spin liquid in the pure Kitaev model 
also has bond-directional and extremely short-range 
spin correlations~\cite{Kitaev2006}. The similar features of spin 
correlations in this QSL state may be owing to the predominant 
Kitaev interaction in our model.

Overall, under the intermediate fields between 
$h_{[111]}^{c1}$ and $h_{[111]}^{c2}$, 
and below the low-temperature scale $T''_{\rm l}$, 
our model calculations predict the presence of the
long-been-sought QSL phase in the compound \RuCl. \\

\noindent{\bf{Discussion}}\\
Firstly, we discuss the nature of the QSL driven by
the out-of-plane fields. In Fig.~\ref{Fig:PD111}d,
the ED calculation suggests a gapless spectrum
in the intermediate phase, and the DMRG simulations
on long cylinders find the logarithmic correction 
in the entanglement entropy scaling (see Supplementary Note~\B{4}),
which further supports a gapless QSL phase. On the other hand, 
the VMC calculations identify the intermediate 
phase as an Abelian chiral spin liquid, which is
topologically nontrivial with a quantized Chern number $\nu = 2$. 
Overall, various approaches consistently find the same 
scenario of two QPTs and an intermediate-field QSL 
phase under high magnetic fields [cf. Fig.~\ref{Fig:PhsDgr}d, f]. 
{It is worth noticing that a similar scenario of a gapless 
QSL phase induced by out-of-plane fields has also been 
revealed in a Kitaev-Heisenberg model with $K > 0$ and 
$J<0$~\cite{Jiang2019}, where the intermediate QSL was found to 
be smoothly connected to the field-induced spin liquid in 
a pure AF Kitaev model~\cite{Zhu2018,Trebst2019,Patel2019PNAS}.
We note the extended AF Kitaev model in Ref.~\cite{Jiang2019} 
can be transformed into a $K$-$J$-$\Gamma$-$\Gamma'$ model 
with FM Kitaev term (and other non-Kitaev terms still distinct 
from our model) through a global spin rotation~\cite{Chaloupka2015}.
Therefore, despite the rather different spin structure factors of the 
field-induced QSL in the AF Kitaev model from ours, it is interesting
to explore the possible connections between the two in the future. 
}

Based on our \RuCl model and precise many-body 
calculations, we offer concrete experimental 
proposals for detecting the intermediate QSL phase 
via the magnetothermodynamic measurements 
under high magnetic fields. The two QPTs are within the 
scope of contemporary technique of pulsed high fields, 
and can be confirmed by measuring the magnetization 
curves~\cite{Zhou2020particlehole,Zhou2020private}.
The specific heat measurements can also be 
employed to confirm the two-transition scenario and 
the high-field gapless QSL states. As field increases, 
the lower temperature scale $T_{\rm l}$ first decreases 
to zero as the zigzag order is suppressed, 
and then rises up again (i.e., $T''_{\rm l}$) in the QSL phase 
[cf. Fig.~\ref{Fig:PD111}a]. As the specific heat exhibits 
a double-peak structure in the high-field QSL regime, 
the thermal entropy correspondingly undergoes a two-step 
release in the QSL phase and exhibits a quasi-plateau near 
the fractional entropy $(\ln{2})/2$ 
[cf. 45.5~T and 91~T lines in Fig.~\ref{Fig:CvS111}b]. 
This, together with the low-$T$ (below $T''_{\rm l}$) algebraic 
specific heat behavior reflecting the gapless excitations, 
can be probed through high-field calorimetry
\cite{Imajo2020highresolution}.

Lastly, it is interesting to note that the emergent high-field 
QSL under out-of-plane fields may be closely related
to the off-diagonal $\Gamma$ term (see, e.g., Refs.~\cite{Gordon2019,Lee2020}) in the compound \RuCl. 
The $\Gamma$ term has relatively small influences in $\textbf{ab}$ plane, 
while introduces strong effects along the $\textbf{c}^*$ axis 
--- from which the magnetic anisotropy in \RuCl mainly 
originates. The zigzag order can be suppressed 
by relatively small in-plane fields and the system 
enters the polarized phase, as the $\Gamma$ 
term does not provide a strong ``protection" of both zigzag and QSL 
phases under in-plane fields (recall the QSL phase in pure FM 
Kitaev model is fragile under external fields~\cite{Jiang2011,Trebst2019,Han2020}).
The situation is very different for out-of-plane fields, where the 
Kitaev (and also Gamma) interactions survive the QSL after the zigzag 
order is suppressed by high fields. Intuitively, the emergence of this QSL 
phase can therefore be ascribed to the strong competition between 
the $\Gamma$ interaction and magnetic field along the hard axis $\textbf{c}^*$ of \RuCl. 
With such insight, we expect a smaller critical field for compounds with a less significant $\Gamma$ interaction. 
In the fast-moving Kitaev materials studies, such compounds 
with relatively weaker magnetic anisotropy, e.g., the recent Na$_2$Co$_2$TeO$_6$ and Na$_2$Co$_2$SbO$_6$
\cite{NCTO2020,Hong2021arXiv,Lin2020arXiv,Songvilay2020}, 
have been found, which may also host QSL induced by 
out-of-plane fields at lower field strengths.
\\

\noindent{\bf{Methods}}\\
\textbf{Exponential tensor renormalization group.} 
The thermodynamic quantities including the specific heat,
magnetic susceptibility, $Z_2$ flux, and the 
spin correlations can be computed with the exponential tensor
renormalization group (XTRG) method~\cite{Chen2018,Lih2019} 
on the Y-type cylinders with width $W=4$ and length up to 
$L=6$ (i.e., YC4$\times$6$\times$2). 
We retain up to $D=800$ states in the XTRG calculations, 
with truncation errors $\epsilon \lesssim 2\times 10^{-5}$,
which guarantees a high accuracy of computed thermal
data down to the lowest temperature $T\simeq1.9$~K.
Note the truncation errors in XTRG, different from 
that in DMRG, directly reflects the relative errors in the
free energy and other thermodynamics quantities.
The low-$T$ data are shown to approach the $T=0$ DMRG 
results (see Fig.~\ref{Fig:PD111}f). In the thermodynamics 
simulations of the \RuCl model, one needs to cover 
a rather wide range of temperatures as the high- and 
low-$T$ scales are different by more than 
one order of magnitude (100~K vs. 7~K in \RuCl 
under zero field). In the XTRG cooling procedure, 
we represent the initial density matrix $\rho_0(\tau)$ at 
a very high temperature $T\equiv 1/\tau$ (with $\tau\ll 1$) 
as a matrix product operator (MPO), and the series of lower
temperature density matrices $\rho_n(2^n \tau)$ ($n\geq1$) 
are obtained by successively multiplying and compressing 
$\rho_n = \rho_{n-1} \cdot \rho_{n-1}$ via the tensor-network 
techniques. Thus XTRG is very suitable to deal with 
such Kitaev model problems, as it cools down 
the system exponentially fast in temperature~\cite{Han2020}.\\

\noindent{\textbf{Density matrix renormalization group.}}
The ground state properties are computed by the density matrix 
renormalization group (DMRG) method, which can be considered as 
a variational algorithm based on the matrix product state (MPS) ansatz. 
We keep up to $D=2048$ states to reduce the truncation errors 
$\epsilon\lesssim 1\times 10^{-8}$ with a very good convergence.
The simulations are based on the high-performance MPS 
algorithm library GraceQ/MPS2~\cite{GQMPS2019}.\\

\noindent{\textbf{Variational Monte Carlo.}}
The ground state of $\alpha$-RuCl$_3$ model are 
evaluated by the variational Monte Carlo (VMC) method 
based on the fermionic spinon representation. 
The spin operators are written 
in quadratic forms of fermionic spinons $S_i^m =\frac{1}{2} C_i^\dagger \sigma^m C_i, m=x,y,z$ 
under the local constraint $\hat{N}_i=c_{i\uparrow}^\dagger c_{i\uparrow} + c_{i\downarrow}^\dagger c_{i\downarrow} = 1$, 
where $C_i^\dagger = (c_{i\uparrow}^\dagger, c_{i\downarrow}^\dagger)$ and $\sigma^m$ are Pauli matrices. 
Through this mapping, the spin interactions are expressed in terms of fermionic operators 
and are further decoupled into a non-interacting mean-field Hamiltonian $H_{\rm mf}(\pmb R)$, 
where $\pmb R$ denotes a set of parameters (see Supplementary Note~\B{5}).  
Then we perform Gutzwiller projection to the mean-field ground state $|\Phi_{\rm mf} (\pmb R)\rangle$ 
to enforce the particle number constraint ($\hat{N}_i= 1$). 
The projected states $|\Psi (\pmb R)\rangle = P_G |\Phi_{\rm mf}(\pmb R) \rangle=
\sum_{\alpha}f(\alpha)|\alpha \rangle$ (here $\alpha$ 
stands for the Ising bases in the many-body Hilbert space, 
same for $\beta$ and $\gamma$ below) provide a series 
of trial wave functions, depending on the specific choice 
of the mean-field Hamiltonian $H_{\rm mf}(\pmb R)$. 
Owing to the huge size of the many-body Hilbert space, 
the energy of the trial state $E (\pmb R) = \langle \Psi(\pmb R) | H | 
\Psi(\pmb R) \rangle / \langle \Psi(\pmb R)| \Psi(\pmb R) \rangle=
\sum_{\alpha}{|f(\alpha)|^2\over\sum_\gamma |f(\gamma)|^2} \Big(\sum_\beta \langle\beta|H|
\alpha\rangle {f(\beta)^*\over f(\alpha)^*}\Big)$ 
is computed using Monte Carlo sampling. 
The optimal parameters $\pmb R$ are determined 
by minimizing the energy $E(\pmb R)$. 
While the VMC calculations are performed on a relatively small size (up to $128$ sites), 
once the optimal parameters are determined we can plot the spinon dispersion of a QSL state 
by diagonalizing the mean-field Hamiltonian on a larger lattice size, e.g., $120\times120$ unit cells in practice.\\

\noindent{\textbf{Exact diagonalization.}}
The 24-site exact diagonalization (ED) is employed 
to compute the zero-temperature dynamical correlations 
and energy spectra. The clusters with periodic boundary 
conditions are depicted in {the inset of 
Fig.~\ref{Fig:Fittings}d} and the Supplementary Information, 
and the \RuCl model under 
in-plane fields ($H_{[11\bar{2}]} \parallel \textbf{a}$ 
and $H_{[1\bar{1}0]} \parallel \textbf{b}$) as well as 
out-of-plane fields $H_{[111]} \parallel \textbf{c}^*$ 
have been calculated, as shown in \Fig{Fig:PD111} 
and Supplementary Note~\B{3}. Regarding the dynamical 
results --- the neutron scattering intensity --- is defined as
\begin{equation}
\begin{split}
\mathcal{I}(\textbf{k}, \omega)\varpropto& f^2(\textbf{k})\int dt\sum_{\mu,\nu}(\delta_{\mu,\nu}-k_{\mu}k_{\nu}/k^2)\\
& \times\sum_{i,j}\langle S_i^{\mu}(t)S_j^{\nu}(0)\rangle e^{i\textbf{k}\cdot(\textbf{r}_j - \textbf{r}_i)} e^{-i\omega t},\\
\end{split}
\label{Intensity}
\end{equation}
where $f(\textbf{k})$ is the atomic form factor of Ru$^{3+}$, 
which can be fitted by an analytical function as reported in 
Ref.~\cite{Cromer1965}. $S^{\mu \nu}(\textbf{k}, \omega)=
\sum_{i,j}\langle S_i^{\mu}(t)S_j^{\nu}(0)\rangle 
e^{i\textbf{k}\cdot(\textbf{r}_j - \textbf{r}_i)} e^{-i\omega t}$
is the dynamical spin structure factor, which can be expressed 
by the continued fraction expansion in the tridiagonal basis 
of the Hamiltonian using Lanczos iterative method. 
For the diagonal part,
\begin{equation}
\begin{split}
S^{\mu\mu}(\textbf{k}, \omega) & = -\frac{1}{\pi}{\rm Im}{\Big [}\langle \psi_0 | \hat{S}^{\mu}_{-\textbf{k}} \frac{1}{z-\hat{H}} \hat{S}^{\mu}_{\textbf{k}} | \psi_0 \rangle {\Big ]}\\
& = -\frac{1}{\pi}{\rm Im} {\Bigg [}  \frac{\langle \psi_0 |\hat{S}^{\mu}_{-\textbf{k}} 
\hat{S}^{\mu}_{\textbf{k}} | \psi_0 \rangle }{z-a_0-\frac{b_1^2}{ z-a_1 - \frac{b_2^2}{z-a_2-\cdot \cdot \cdot}}} {\Bigg]},\\
\end{split}
\label{Intensity}
\end{equation}
\\
where $z = \omega +E_0 +i \eta$, 
$E_0$ is the ground state energy, $| \psi_0 \rangle$ is the 
ground state wave function, $\eta$ is the Lorentzian broadening 
factor (here we take $\eta=0.5$ meV in the calculations, 
i.e., 0.02 times the Kitaev interaction strength $|K|=$25~meV),
and $a_i$ ($b_{i+1}$) is the diagonal 
(sub-diagonal) matrix element
of the tridiagonal Hamiltonian.
On the other hand,
for the off-diagonal part,
we define a Hermitian operator 
$\hat{S}_{\textbf{k}}^{\mu} + \hat{S}_{\textbf{k}}^{\nu}$
to do the continued fraction expansion,
\begin{equation}
\begin{split}
&S^{(\mu+\nu)(\mu+\nu)}(\textbf{k}, \omega) \\
&\hspace{3mm} = -\frac{1}{\pi}{\rm Im}{\Big [}\langle \psi_0 | (\hat{S}^{\mu}_{-\textbf{k}} + \hat{S}^{\nu}_{-\textbf{k}}) \frac{1}{z-\hat{H}} (\hat{S}^{\mu}_{\textbf{k}}+\hat{S}^{\nu}_{\textbf{k}}) | \psi_0 \rangle {\Big ]}\\
&\hspace{3mm} = -\frac{1}{\pi}{\rm Im} {\Bigg [}  \frac{\langle \psi_0 |(\hat{S}^{\mu}_{-\textbf{k}} + \hat{S}^{\nu}_{-\textbf{k}}) (\hat{S}^{\mu}_{\textbf{k}}+\hat{S}^{\nu}_{\textbf{k}}) | \psi_0 \rangle }{z-a_0-\frac{b_1^2}{ z-a_1 - \frac{b_2^2}{z-a_2-\cdot \cdot \cdot}}} {\Bigg]}.\\
\end{split}
\label{Intensity}
\end{equation}
Then the off-diagonal $S^{\mu \nu}(\textbf{k}, \omega)$
can be computed by 
$S^{\mu\nu}(\textbf{k}, \omega) + S^{\nu \mu}(\textbf{k}, \omega) = 
S^{(\mu+\nu)(\mu+\nu)}(\textbf{k}, \omega) - S^{\mu \mu}(\textbf{k}, \omega) 
- S^{\nu \nu}(\textbf{k}, \omega)$.
Following the INS experiments~\cite{Banerjee2017}, 
the shown scattering intensities in Fig.~\ref{Fig:Fittings}
are integrated over perpendicular momenta $k_z \in [-5 \pi, 5 \pi]$,
assuming perfect two-dimensionality of \RuCl in the ED calculations. \\

\noindent{\bf{Data availability}}\\
The data that support the findings of this study are 
available from the corresponding author upon reasonable request.\\

\noindent{\bf{Code availability}}\\
All numerical codes in this paper are available 
upon request to the authors. \\

\bibliography{kitaevRef} 

$\,$\\
\textbf{Acknowledgements} \\
We are indebted to Yang Qi, Xu-Guang Zhou, Yasuhiro Matsuda, 
Wentao Jin, Weiqiang Yu, Jinsheng Wen, Kejing Ran, 
Yanyan Shangguan and Hong Yao for helpful discussions. 
This work was supported by the National Natural Science 
Foundation of China (Grant Nos. 11834014, 11974036, 11974421, 11804401),
Ministry of Science and Technology of China (Grant No. 2016YFA0300504), 
and the Fundamental Research Funds for the Central Universities 
(BeihangU-ZG216S2113, SYSU-2021qntd27). We thank the 
High-Performance Computing Cluster of Institute of Theoretical 
Physics-CAS for the technical support and generous allocation 
of CPU time.

$\,$\\
\textbf{Author contributions} \\
W.L., S.S.G. and Z.X.L initiated this work. 
H.L. and D.W.Q. performed the thermal tensor network calculations, 
H.L. obtained the Hamiltonian parameters by fitting experiments, 
Y.G. confirmed the model parameters with Bayesian searching, 
H.K.Z. (DMRG) and J.C.W. (VMC) performed the ground state simulations,
and H.Q.W. computed the dynamical correlations by ED.
All authors contributed to the analysis of the results and the preparation 
of the draft. S.S.G., Z.X.L., and W.L. supervised the project. 

$\,$\\
\textbf{Competing interests} \\
The authors declare no competing interests. 

$\,$\\
\textbf{Additional information} \\
\textbf{Supplementary Information} is available in the online version of the paper. \\
\noindent

\newpage
\onecolumngrid
\appendix

\setcounter{equation}{0}
\setcounter{figure}{0}
\setcounter{table}{0}
\setcounter{subsection}{0}

\renewcommand{\thesubsection}{\normalsize{Supplementary Note \arabic{subsection}}}
\renewcommand{\theequation}{S\arabic{equation}}
\renewcommand{\thefigure}{\arabic{figure}}
\renewcommand{\thetable}{\arabic{table}}

\subsection{Determination of the Effective $\alpha$-RuCl$_3$ Hamiltonian}
\label{SN:SHFitting}
\textbf{Fitting the model parameters from thermodynamic 
{measurements.}} {In this section we show the workflow of the 
model parameter fittings. By performing exponential tensor 
renormalization (XTRG) calculations on a YC$4\times$4$\times$2
lattice, we scan the parameter space spanned by the couplings  
$[K, \Gamma, \Gamma', J]$, and fit the specific 
heat $C_{\rm{m}}(T)$, in-plane ($\chi_{\rm ab}$) 
and out-of-plane susceptibilities ($\chi_{\rm c^*}$) measured 
under a small magnetic field $\mu_0H\simeq 1$~T.
From the susceptibility simulations, we also determine 
the corresponding $g$-factors, $g_{\rm ab}$ and $g_{\rm c^*}$, 
along with other couplings. Given the determined parameters, 
we can compute the static spin structure factors, 
and confirm the appearance of zigzag magnetic order 
at low temperature (see Figs.~\B{2,3} in the main text). 
XTRG and density matrix renormalization group 
(DMRG) are employed to compute the magnetization 
curve and compared to experiments directly. 
Exact diagonalization (ED) calculations are performed 
on 24-site clusters \rev{[see Supplementary \Fig{Fig:Q3ED}(a,b) insets],}
from which we obtain the dynamical spin structure factors.
}

\begin{figure}[h!]
\includegraphics[angle=0,width=1\linewidth]{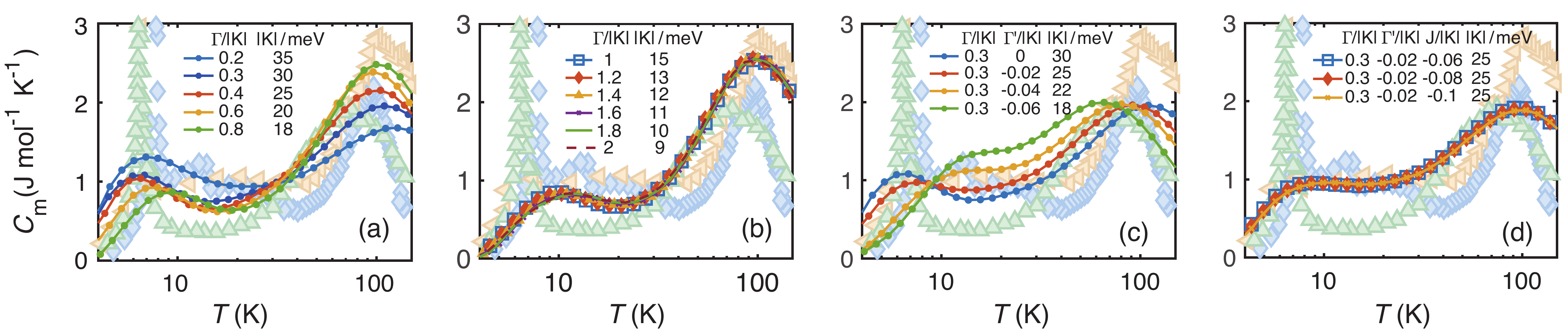}
\renewcommand{\figurename}{\textbf{Supplementary Figure}}
\caption{\textbf{Simulations of the magnetic specific heat 
$C_{\rm m}$.} (a,b) show the $C_{\rm m}$ curves with different 
$\Gamma$ terms, with rest interactions temporarily 
set to be zero. (c) shows the simulated $C_{\rm m}$ with
various $\Gamma'$ couplings and fixed $\Gamma/|K|=0.3$, 
and (d) checks the effects of Heisenberg term $J$ with fixed 
$\Gamma=0.3$ and $\Gamma'=-0.02$. All those curves are 
compared with experimental data \cite{Kubota2015,Do2017,Widmann2019}, 
which are shown in the background. The energy scale $|K|$ 
is tuned adaptively in different cases to fit the $C_{\rm m}$ curves.}
\label{FigS:FitCm}
\end{figure}

\begin{figure}[h!]
\includegraphics[angle=0,width=1\linewidth]{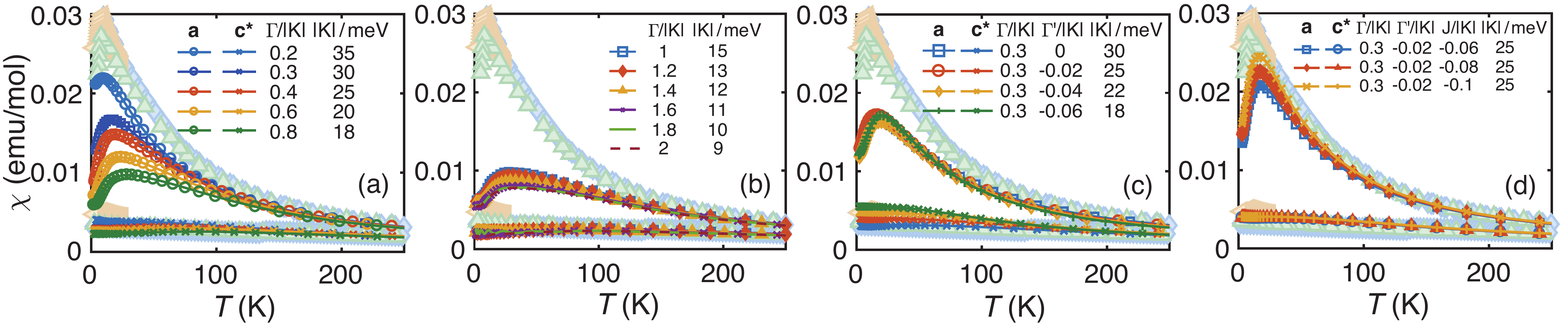}
\renewcommand{\figurename}{\textbf{Supplementary Figure}}
\caption{\textbf{Simulations of the magnetic susceptibility 
$\chi_{\rm ab}$ and $\chi_{\rm c^*}$.} (a,b) The susceptibility 
results, including the in-plane susceptibility $\chi_{\rm ab}$ 
computed under fields $H_{[11\bar{2}]} \parallel \textbf{a}$
and the out-of-plane one $\chi_{\rm c^*}$ 
under $H_{[111]}\parallel \textbf{c}^*$, 
are shown with different $\Gamma/|K|$ values.
(c) shows the fittings 
with various $\Gamma'$ and a fixed $\Gamma=0.3$, 
and (d) the various Heisenberg $J$ with 
fixed $\Gamma=0.3$ and $\Gamma'=-0.02$.
The two $g$-factors, i.e., $g_{\rm ab}$ and $g_{\rm c^*}$, 
are determined to be between 2.2-2.5 
(depending on other coupling parameters) 
from fitting the high-temperature susceptibility. 
The magnetic field $\mu_0H$ involved
in our fittings are between 0.8-1~T,
in consistent with the experimental field of 1~T.}
\label{FigS:FitChi}
\end{figure}

To be concrete, we show in Supplementary 
Figs.~\ref{FigS:FitCm},~\ref{FigS:FitChi}
part of our simulated data in the thermodynamic properties. 
In Supplementary Fig.~\ref{FigS:FitCm}(a,b), we start with scanning over 
various $\Gamma$ values \rev{with ferromagnetic $K<0$}, 
by setting $\Gamma'=J=0$ at first. As shown in Supplementary Fig.~\ref{FigS:FitCm}(a), 
the $C_{\rm{m}}$ curves are sensitive to 
$\Gamma$ for $\Gamma < 1$, while the curves 
do not change much for $\Gamma \geq 1$
in Supplementary Fig.~\ref{FigS:FitCm}(b), given that the energy 
scale $|K|$ is properly tuned. Therefore, to 
uniquely pinpoint the parameter $\Gamma/|K|$, 
we need to include more thermodynamic 
measurements like the magnetic susceptibility.

As shown in Supplementary Fig.~\ref{FigS:FitChi}(a,b), as the $\Gamma$ 
interaction increases, the height of computed susceptibility 
peak decreases accordingly and deviates the experimental 
in-plane susceptibility $\chi_{\rm ab}$ curves. 
After a thorough scanning, we find $\Gamma=0.3$ 
constitutes an overall optimal parameter in our 
model fittings of the specific heat and susceptibility. 
With the given $\Gamma$, we find in Supplementary Fig.~\ref{FigS:FitChi}(c)  
the susceptibility curves turn out to be not very sensitive 
to the {small} $\Gamma'$ interactions, while,
on the other hand, the low-$T$ peak of $C_{\rm m}$ moves 
towards higher temperatures as $|\Gamma'|$ increases in 
Supplementary Fig.~\ref{FigS:FitCm}(c). This can be understood as the 
$\Gamma'$ term is crucial for stabilizing the zigzag 
order {when $K<0$} and thus has strong influences
on the low-$T$ peak. Therefore, we fix $\Gamma'=-0.02$ 
from the specific heat fittings.

Lastly, we determine the nearest-neighboring Heisenberg 
term $J$. In Supplementary Fig.~\ref{FigS:FitChi}(d), we find the height of the
susceptibility $\chi_{\rm ab}$ peak enhances and thus 
approaches the experimental curves as $J$ changes 
from $-0.06$ to $-0.1$, while the specific heat $C_{\rm m}$
is not sensitive to $J$, as shown in Supplementary Fig.~\ref{FigS:FitCm}(d). 
From these careful scanings in the parameter space, 
we determine the parameter set as $[\Gamma/|K|, 
\Gamma'/|K|, J/|K|, g_{\rm ab}, g_{\rm c^*}] = [0.3, -0.02, -0.1, 2.5, 2.3]$ 
with $K=-25$ meV, which can very well fit both the specific heat 
as well as the in- and out-of-plane magnetic susceptibilities.

\begin{figure}[h!]
\includegraphics[angle=0,width=1\linewidth]{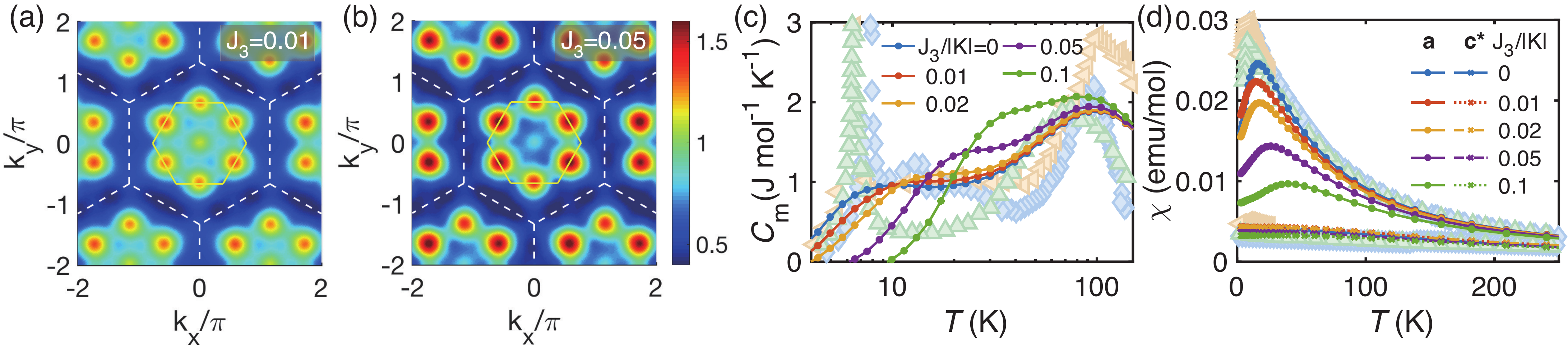}
\renewcommand{\figurename}{\textbf{Supplementary Figure}}
\caption{\textbf{Effects of the $J_3$ coupling 
in magnetic structure and thermodynamics.} (a,b) show the 
static spin structure factors of our \RuCl model 
($\Gamma/|K|=0.3$, $\Gamma'/|K|=-0.02$, $J/|K|=-0.1$) 
with additional $J_3=$ 0.01 and 0.05, respectively. 
The results are computed at $T\simeq2.7$~K, 
and the increased M-point intensity with $J_3$ 
shows that the latter enhances the zigzag order.
(c) shows the specific heat $C_{\rm{m}}$ results 
with various $J_3>0$ interactions, as compared 
to the experimental data in the background. 
(d) The in- and out-of-plane susceptibility results
of our \RuCl model with additional $J_3$ values.}
\label{FigS:ParaJ3}
\end{figure}

{\textbf{The third-nearest neighboring $J_3$ interaction.}}
Besides the parameter set considered above,
a third neighboring Heisenberg interaction 
$J_3$ has also been suggested to stabilize the zigzag order 
\rev{for $K<0$}~\cite{Winter2017,Winter2017NC,Winter2018},
which can play a similar role to the off-diagonal $\Gamma'$ 
interaction. In order to explore its effects, we introduce 
an additional $J_3$ term to our \RuCl model and show 
the computed results in Supplementary Fig.~\ref{FigS:ParaJ3}. 

In the spin structure factors in Supplementary Fig.~\ref{FigS:ParaJ3}(a,b), 
we find the M-point intensity clearly enhances when
$J_3$ increases from 0.01 to 0.05. At the same time,
even a small $J_3$ can have a considerable impact 
on the low-temperature thermodynamics and makes 
our model fittings deviate from the experimental data. 
In Supplementary Fig.~\ref{FigS:ParaJ3}(c,d), we find a $J_3 \geq 0.05$ 
clearly spoils the fittings to $C_{\rm m}$ and $\chi$: 
The low-$T$ specific heat peak moves towards higher 
temperature, and the height of the susceptibility peak 
gets reduced when increasing $J_3$. In addition,
from Supplementary Fig.~\ref{FigS:ParaJ3}(b), $J_3=0.05$ 
interaction evidently cripples the spin intensity
at the $\Gamma$ point of the Brillouin zone (BZ).
Overall, we find $J_3$ indeed plays a very similar role 
as $\Gamma'$ in stabilizing the zigzag order, and, 
such a term, if exists, should have a small
coupling strength. We therefore leave out $J_3$ 
in the main text as well as the discussion below, 
and mostly focus on the minimal $K$-$J$-$\Gamma$-$\Gamma'$ 
effective model in the main text.

\begin{figure}[h!]
\includegraphics[angle=0,width=0.8\linewidth]{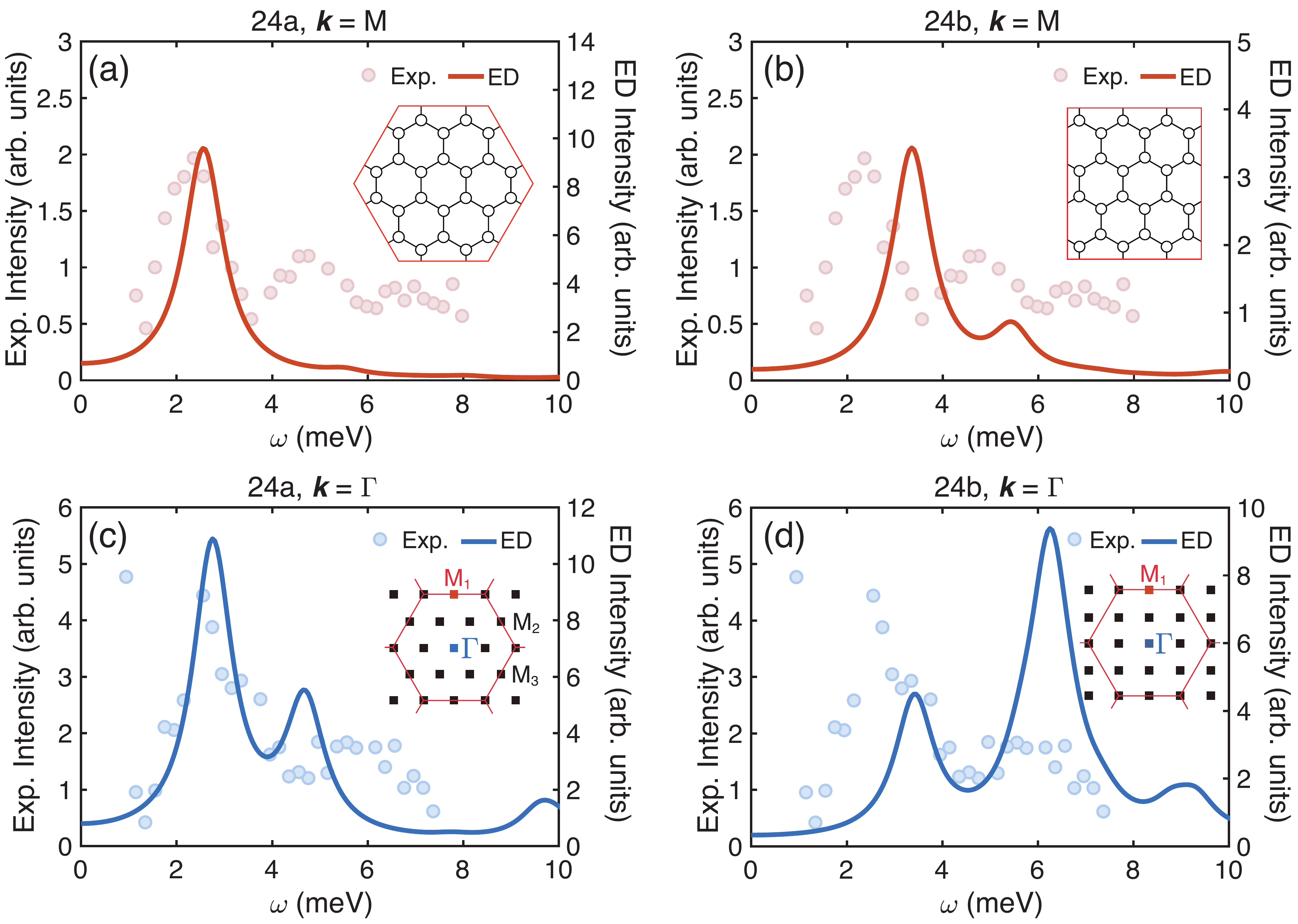}
\renewcommand{\figurename}{\textbf{Supplementary Figure}}
\caption{\rev{\textbf{The dynamical ED results of intensity 
$\mathcal{I}(\textbf{k}, \omega)$ at the K and $\Gamma$ points.}
(a,b) shows the constant $\textbf{k}$-cut of $\mathcal{I}(\textbf{k}, \omega)$
at \textbf{k} = M and (c,d) at \textbf{k} = $\Gamma$ points.
The dynamical data in (a,c) are computed on cluster 24a with 
its real-space and reciprocal lattices illustrated in panels (a) and (c), 
respectively. The corresponding results of 24b cluster are 
shown in (b,d), note there are no M$_2$ and M$_3$ points 
in the BZ of 24b cluster as illustrated in the inset of (d). The intensity 
peak locations $\omega_{\rm M}$ and $\omega_\Gamma$ 
are stable [$\omega_{\rm M}$ = 2.5 and 3.3~meV in (a,b), 
and $\omega_\Gamma$ =  2.7 and 3.4~meV in (c,d), respectively] 
while the relative intensities between two peaks are found more 
sensitive to the specific geometries due to the finite-size effects.}
}
\label{Fig:Q3ED}
\end{figure}

\rev{\textbf{Exact diagonalization results of dynamical spin structure factors.}
In the dynamical simulations, the 24-site cluster with $C_3$ symmetry 
{(denoted as 24a henceforth)} has been widely adopted in ED calculations of the 
Kitaev model (see, e.g., Refs.~\cite{Winter2017NC,Gordon2019,Trebst2019,
Laurell2020}). Amongst other geometries that are accessible by ED, 
the 24a cluster is unique as it contains all the high-symmetry points
in the BZ [c.f., inset in Supplementary \Fig{Fig:Q3ED}(c)], which is important for dynamical 
property calculations. Therefore, we choose the 24a cluster in presenting 
our dynamical data in the main text. Nevertheless, in Supplementary \Fig{Fig:Q3ED} 
we also perform dynamical ED simulations on a different 24-site 
geometry (24b) and compare the results to those of the 24a cluster. 
It can be seen in Supplementary \Fig{Fig:Q3ED} that the positions $\omega_{\rm M}$ 
and $\omega_\Gamma$ of the intensity peaks as well as the double-peak 
structure of $\Gamma$-intensity curves are qualitatively consistent. 
Note the 24b cluster shown in Supplementary \Fig{Fig:Q3ED}(b,d) are not $C_3$ 
symmetric and thus does not possess all high symmetry moment points.}

\textbf{Magnetic anisotropy and the off-diagonal 
$\Gamma$ term.} The off-diagonal $\Gamma$ 
interaction has been suggested to be responsible
for the strong magnetic anisotropy between
in- and out-of-plane directions~\cite{Lampen-Kelley2018,Sears2020}. 
In a pure Kitaev model without the $\Gamma$ term, 
it has been shown that the in-plane ($\chi_{\rm ab}$) 
and out-of-plane ($\chi_{\rm c^*}$) susceptibilities are 
of similar magnitudes~\cite{Han2020}. On the other hand, 
as shown in Supplementary Fig.~\ref{FigS:FitChi}, when the 
$\Gamma$ term is introduced, we find the two 
susceptibility curves are clearly separated,
suggesting that the off-diagonal $\Gamma$ interaction 
indeed constitutes a resource of strong easy-plane 
anisotropy observed in \RuCl.

\textbf{Automatic searching of the Hamiltonian parameters.}
Through the fitting process described above, we have manually 
determined an accurate set of model parameters of \RuCl.
As a double-check, below we exploit the recently developed 
automatic parameter searching technique based on the 
Bayesian approach~\cite{Yu2020Learning} to conduct 
a global optimization of the model parameters in Supplementary Fig.~\ref{FigS:LndScp}. 
The target is to minimize the fitting loss of the simulated 
results to measured thermodynamic quantities, 
including the specific heat $C_{\rm m}$ as well as the 
in- ($\chi_{\rm ab}$) and out-of-plane ($\chi_{\rm c^*}$) susceptibilities.
Since the specific heat data vary greatly in various measurements 
and the lower-$T$ peak diverges likely due to 3D effects, 
we take the mean values of several measurements as the 
specific heat data $C_{\rm m}^{\rm exp}(T)$ used in the fittings. 
\rev{In practice, the loss function $\mathcal{L}$ is designed as follows: 
\begin{equation}
\begin{split}
\mathcal{L} = \text{log} \{ \frac{\lambda_\chi}{N_{\chi_{\rm ab}}}
\sum_{T>T_{\rm cut}}[{\chi_{\rm ab}^{\rm exp}(T)-\chi_{\rm ab}^{\rm sim}(T)}]^2 
& +\frac{\lambda_\chi}{N_{\chi_{\rm c^*}}}\sum_{T>T_{\rm cut}}
[{\chi_{\rm c^*}^{\rm exp}(T)-\chi_{\rm c^*}^{\rm sim}(T)}]^2 \\
& +\frac{\mathcal{P} \, \lambda_C}{N_{C_{\rm m}}}\sum_{T>T'_{\rm cut}}
[{C_{\rm m}^{\rm exp}(T)-C_{\rm m}^{\rm sim}(T)}]^2  
\},
\end{split}
\end{equation}
where $\lambda_C=1/ {\mathrm{max}(C_{\rm m}^{\rm exp})}^2$ and 
$\lambda_\chi=1/{\mathrm{max}(\chi_{ab}^{\rm exp})}^2$ controls the 
weights of specific heat and magnetic susceptibility in defining the fitting 
loss function. $N_{\chi_{\rm ab}}$, $N_{\chi_{\rm c^*}}$, and $N_{C_{\rm m}}$
are the data point numbers in three experimental curves, respectively. 
To focus on the fittings of the locations of two specific heat peaks at 100~K 
and 8~K, respectively, we further introduce a penalty factor $\mathcal{P}$
defined as
\begin{equation}
\mathcal{P}=\left\{
\begin{matrix}
\text{min}[\frac{1}{2}\sum_i^{1,2}\text{exp}({\frac{|T_i^{\rm exp}-T_i^{\rm sim}|}{0.15 \, T_i^{\rm exp}}})-1,a] + 
\text{min}[\frac{1}{2}\sum_i^{1,2}\text{exp}({\frac{|C_i^{\rm exp}-C_i^{\rm sim}|}{0.25 \, C_i^{\rm exp}}})-1,a] &, 
&\text{ two peaks}\\
2a &, & \text{otherwise} 
\end{matrix}
\right.
\end{equation}
where $\{T_1, T_2\} = \{100~\text{K}, 8~\text{K}\}$ are the position 
of the peaks, with $\{C_1, C_2\}$ the specific heat value at corresponding 
higher and lower $T$. As our many-body calculations are performed on a 
finite-size system, we need to introduce the $T_{\rm cut}(T'_{\rm cut})$ as the lowest 
temperature involved in the fittings, and in practice it is set as $T_{\rm cut}=25$~K 
for magnetic susceptibility and $T'_{\rm cut}=3$~K for the specific heat data. 
The factor $\mathcal{P}$ emphasizes the double-peak structure 
as well as the locations and {heights} of each peaks in the simulated 
$C_{\rm m}$, where we set an empirical factor $a = 2$ to enlarge the 
loss function as a penalty for the simulated specific heat curves without 
double-peak structure.} 

\begin{figure}[h!]
\includegraphics[angle=0,width=\linewidth]{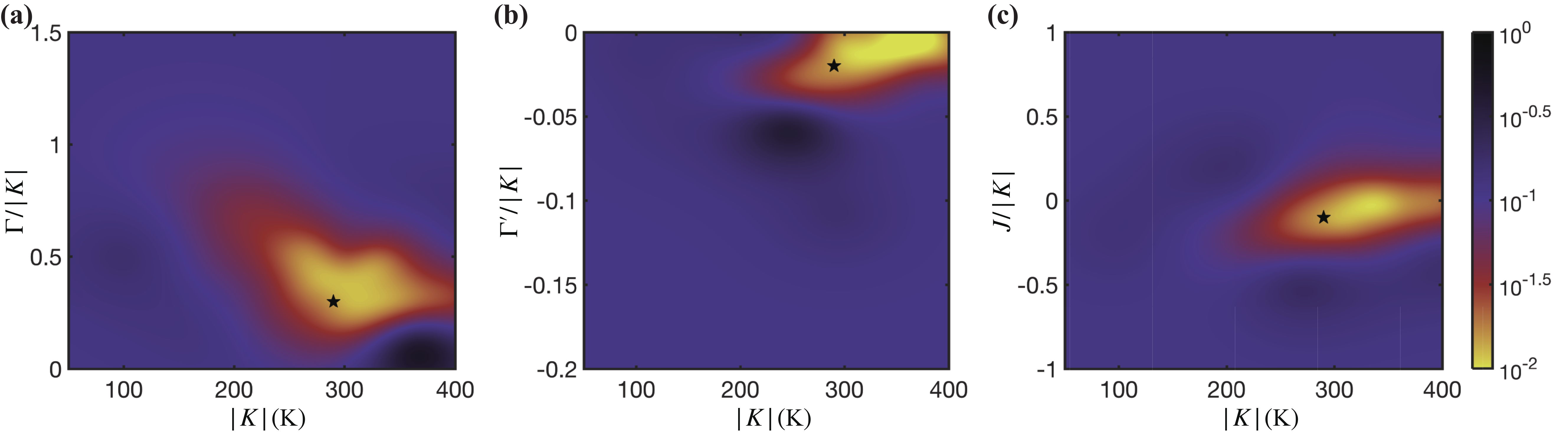}
\renewcommand{\figurename}{\textbf{Supplementary Figure}}
\caption{\rev{{\textbf{The estimated fitting loss $\mathcal{L}$ shown in (a) $K$-$\Gamma$, 
(b) $K$-$\Gamma'$, and (c) $K$-$J$ planes}. The results are} 
obtained by Bayesian optimization through 120 iterations of 
XTRG calculations on the YC4$\times$4$\times$2 cylinder. 
The asterisk represents the optimal parameter set found to also 
reproduce other major experimental observations including the 
magnetization curve and dynamical spin structures, which is located 
within the bright regime with small fitting loss of thermodynamics. 
The two Land\'e factors $g_{\rm ab}$ and $g_{\rm c^*}$ 
at each parameter point are tuned to their optimal values 
(between 1.5 and 3) to minimize the loss $\mathcal{L}$. Note, 
when plotting the cross sectors, like the $K$-$\Gamma$ plane 
in (a), the other model parameters are fixed at their optimal values.
}}
\label{FigS:LndScp}
\end{figure}
In Supplementary Fig.~\ref{FigS:LndScp}, we show the estimated 
loss $\mathcal{L}$ in the $K$-$\Gamma$, 
$K$-$\Gamma'$, and $K$-$J$ planes, respectively. 
From the color map of $\mathcal{L}$, 
we see clearly optimal (bright) regimes, 
and the determined optimal parameter set in the main text
is indicated by the asterisk, i.e., $K=-25$ meV, 
$\Gamma=0.3\ |K|$, $\Gamma'=-0.02\ |K|$, and $J=-0.1\ |K|$,
with the in- and out-of-plane Land\'e factors found to be 
$g_{\rm ab} = 2.5$ and $g_{\rm c^*} = 2.3$, respectively,
which can fit excellently both the thermodynamic and 
dynamic measurements in Fig.~\B{2} of the main text.

\subsection{Revisit of Various  \RuCl Candidate Models}
\label{SN:RevCandMod}
We now employ XTRG to revisit some of the previously proposed 
\RuCl candidate models in the literature~\cite{Winter2016,Winter2017,
Wu2018,Cookmeyer2018,Kim2016,Suzuki2019,Ran2017,Wang2017,
Ozel2019}, where the Kitaev coupling $K$, Heisenberg $J$ 
(nearest neighboring), $J_3$ (third-nearest neighboring), 
as well as off-diagonal $\Gamma$ and $\Gamma'$ terms 
are considered. We calculated the thermodynamics and static 
spin structure factors, the main results are summarized in 
Supplementary Table.~\ref{Tab:npj} and the detailed thermodynamic 
results are shown in Supplementary Figs.~\ref{FigS:Cm-npj} 
and \ref{FigS:SSnpj}. 

\begin{table*}[h]
\begin{tabular}{llllllllllll}
\hline
\multicolumn{1}{|c|}{Refs.} 
& \multicolumn{1}{c|}{\ $K$ (meV)\ } & \multicolumn{1}{c|}{\ $\Gamma/|K|$ \ } & \multicolumn{1}{c|}{\ $\Gamma'/|K|$\ } & \multicolumn{1}{c|}{\ $J/|K|$\ } & \multicolumn{1}{c|}{\ $J_3/|K|$\ } & \multicolumn{1}{c|}{\ \ ${C_{\rm{m}}}^{\dag}$\ \ } & \multicolumn{1}{c|}{\ \ $\chi_{a}$\ \ } & \multicolumn{1}{c|}{\ \ $\chi_{\rm c^*}$\ \ } & \multicolumn{1}{c|}{\ zigzag order$^{\ddag}$\ } & \multicolumn{1}{c|}{\ M-star $^{\ast}$ \ } \\ \hline

\multicolumn{1}{|c|}{Winter2016~\cite{Winter2016}} & \multicolumn{1}{c|}{-6.7}  & \multicolumn{1}{c|}{0.985}  & \multicolumn{1}{c|}{-0.134} &  \multicolumn{1}{c|}{-0.254} & \multicolumn{1}{c|}{0.403}& \multicolumn{1}{c|}{\XSolidBrush}& \multicolumn{1}{c|}{\XSolidBrush} & \multicolumn{1}{c|}{\Checkmark}& \multicolumn{1}{c|}{\Checkmark} & \multicolumn{1}{c|}{\XSolidBrush}  \\

\multicolumn{1}{|c|}{Winter2017~\cite{Winter2017}} & \multicolumn{1}{c|}{-5}  & \multicolumn{1}{c|}{0.5}  & \multicolumn{1}{c|}{/} &  \multicolumn{1}{c|}{-0.1} & \multicolumn{1}{c|}{0.1}& \multicolumn{1}{c|}{{\XSolidBrush}}& \multicolumn{1}{c|}{{\XSolidBrush}}& \multicolumn{1}{c|}{{\XSolidBrush}}& \multicolumn{1}{c|}{\Checkmark} & \multicolumn{1}{c|}{\XSolidBrush} \\

\multicolumn{1}{|c|}{Wu2018~\cite{Wu2018}} & \multicolumn{1}{c|}{-2.8}  & \multicolumn{1}{c|}{0.857}  & \multicolumn{1}{c|}{/} &  \multicolumn{1}{c|}{-0.125} & \multicolumn{1}{c|}{0.121}& \multicolumn{1}{c|}{{\XSolidBrush}}& \multicolumn{1}{c|}{{\XSolidBrush}}& \multicolumn{1}{c|}{{\XSolidBrush}}& \multicolumn{1}{c|}{\Checkmark}  & \multicolumn{1}{c|}{\XSolidBrush} \\

\multicolumn{1}{|c|}{Cookmeyer2018~\cite{Cookmeyer2018}} & \multicolumn{1}{c|}{-5}  & \multicolumn{1}{c|}{0.5}  & \multicolumn{1}{c|}{/} &  \multicolumn{1}{c|}{-0.1} & \multicolumn{1}{c|}{0.023}& \multicolumn{1}{c|}{{\XSolidBrush}}& \multicolumn{1}{c|}{{\XSolidBrush}}& \multicolumn{1}{c|}{{\XSolidBrush}}& \multicolumn{1}{c|}{\Checkmark}& \multicolumn{1}{c|}{\Checkmark} \\

\multicolumn{1}{|c|}{Kim2016~\cite{Kim2016}} & \multicolumn{1}{c|}{-6.55}  & \multicolumn{1}{c|}{0.802}  & \multicolumn{1}{c|}{-0.145} &  \multicolumn{1}{c|}{-0.234} & \multicolumn{1}{c|}{/} & \multicolumn{1}{c|}{{\XSolidBrush}} & \multicolumn{1}{c|}{{\Checkmark}}& \multicolumn{1}{c|}{{\Checkmark}}& \multicolumn{1}{c|}{\Checkmark}& \multicolumn{1}{c|}{\XSolidBrush} \\

\multicolumn{1}{|c|}{Suzuki2019~\cite{Suzuki2019}} & \multicolumn{1}{c|}{-24.4}  & \multicolumn{1}{c|}{0.215}  & \multicolumn{1}{c|}{-0.039} &  \multicolumn{1}{c|}{-0.063} & \multicolumn{1}{c|}{/} & \multicolumn{1}{c|}{{\Checkmark}} & \multicolumn{1}{c|}{{\Checkmark}}& \multicolumn{1}{c|}{{\Checkmark}}& \multicolumn{1}{c|}{\Checkmark}& \multicolumn{1}{c|}{\XSolidBrush}\\

\multicolumn{1}{|c|}{Ran2017~\cite{Ran2017}} & \multicolumn{1}{c|}{-6.8}  & \multicolumn{1}{c|}{1.397}  & \multicolumn{1}{c|}{/} &  \multicolumn{1}{c|}{/} & \multicolumn{1}{c|}{/} & \multicolumn{1}{c|}{{\Checkmark}}& \multicolumn{1}{c|}{{\Checkmark}}& \multicolumn{1}{c|}{{\Checkmark}}& \multicolumn{1}{c|}{{\XSolidBrush}}& \multicolumn{1}{c|}{\Checkmark} \\

\multicolumn{1}{|c|}{Wang2017~\cite{Wang2017}} & \multicolumn{1}{c|}{-10.9}  & \multicolumn{1}{c|}{0.56}  & \multicolumn{1}{c|}{/} &  \multicolumn{1}{c|}{-0.028} & \multicolumn{1}{c|}{0.003} & \multicolumn{1}{c|}{{\XSolidBrush}}& \multicolumn{1}{c|}{{\Checkmark}}& \multicolumn{1}{c|}{{\Checkmark}}& \multicolumn{1}{c|}{{\XSolidBrush}}& \multicolumn{1}{c|}{\XSolidBrush} \\

\multicolumn{1}{|c|}{Ozel2019~\cite{Ozel2019}} & \multicolumn{1}{c|}{-3.5}  & \multicolumn{1}{c|}{0.671}  & \multicolumn{1}{c|}{/} &  \multicolumn{1}{c|}{0.131} & \multicolumn{1}{c|}{/}& \multicolumn{1}{c|}{{\XSolidBrush}}& \multicolumn{1}{c|}{{\XSolidBrush}}& \multicolumn{1}{c|}{{\XSolidBrush}}& \multicolumn{1}{c|}{{\XSolidBrush}}& \multicolumn{1}{c|}{\XSolidBrush} \\

\multicolumn{1}{|c|}{Our model$^{\star}$} & \multicolumn{1}{c|}{-25}  & \multicolumn{1}{c|}{0.3}  & \multicolumn{1}{c|}{-0.02} &  \multicolumn{1}{c|}{-0.1} & \multicolumn{1}{c|}{/} & \multicolumn{1}{c|}{{\Checkmark}}  & \multicolumn{1}{c|}{{\Checkmark}}  & \multicolumn{1}{c|}{{\Checkmark}}  & \multicolumn{1}{c|}{\Checkmark} & \multicolumn{1}{c|}{\Checkmark}  \\[1ex] \hline

\multicolumn{11}{|l|}{${\dag}$~Check if the double-peak feature exists in the calculated $C_{\rm{m}}$ curves.} \\[1ex] \hline

\multicolumn{11}{|l|}{$\ddag$~Check if the model exhibits a low-$T$ zigzag order.} \\[1ex] \hline

\multicolumn{11}{|l|}{$\ast$~Check if the dynamical M-star structure exists (taken from Ref.~\cite{Laurell2020}, except for our model in the last row).} \\[1ex] \hline

\multicolumn{11}{|l|}{$\star$~Our \RuCl model proposed in the main text.} \\[1ex] \hline

\end{tabular}
\renewcommand{\tablename}{\textbf{Supplementary Table}}
\caption{\textbf{Checklist of various candidate models of \RuCl
on typical experimental features.} The ``{\Checkmark} '' 
symbol indicates a good fitting between model calculations 
and experimental measurements, while ``{\XSolidBrush} '' 
represents a disagreement between the two.}
\label{Tab:npj}
\end{table*}

\begin{figure}[h!]
\includegraphics[angle=0,width=0.55\linewidth]{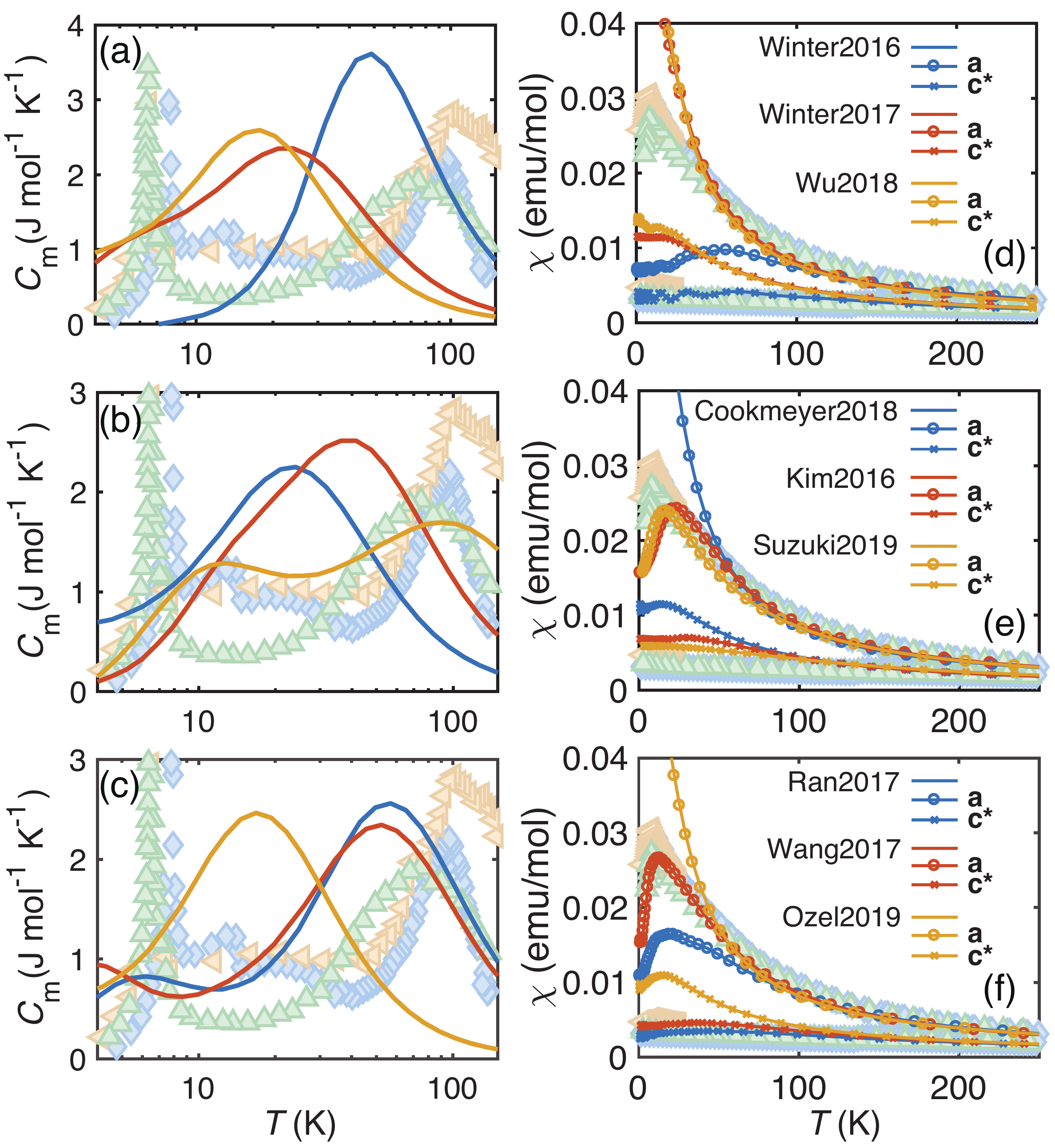}
\renewcommand{\figurename}{\textbf{Supplementary Figure}}
\caption{\textbf{Thermodynamic properties of various \RuCl candidate models.}
(a-c) The magnetic specific heat $C_{\rm m}$ curves (marked as solid lines) 
compared to the experimental measurements. (d-f) The susceptibility data 
with fields $\mu_0H \parallel$ \textbf{a} and \textbf{c$^*$} around 1~T.
The calculations are performed on YC4$\times$4$\times$2 lattice.
}
\label{FigS:Cm-npj}
\end{figure}

\textbf{Specific heat and susceptibility curves.}
In Supplementary Fig.~\ref{FigS:Cm-npj} we show the simulated magnetic specific 
heat $C_{\rm m}$ and susceptibility $\chi$ of various candidate models, 
and compare them to the experimental measurements. From Supplementary Fig.~\ref{FigS:Cm-npj}(a-c), 
we find only the models Suzuki2019~\cite{Suzuki2019} and Ran2017~\cite{Ran2017} 
exhibit a double-peaked $C_{\rm m}$ curve, each located at the characteristic 
temperature precisely as in experiments. For the rest of revisited candidate 
models, however, we did not observe the desired double-peak feature 
in the right temperature window.

The magnetic susceptibility results of the candidate models 
are shown in Supplementary Fig.~\ref{FigS:Cm-npj}(d-f), where we find four 
models out of them, i.e., Kim2016~\cite{Kim2016},
Suzuki2019~\cite{Suzuki2019}, Ran2017~\cite{Ran2017}, 
and Wang2017~\cite{Wang2017} offer adequate fittings 
to both in- and out-of-plane susceptibility measurements, 
with similar Land\'e factors $g_{\rm ab}$ and $g_{\rm c^*}$
ranging between 2.1 and 2.3 (depending on the specific 
candidate model).

\begin{figure}[h!]
\includegraphics[angle=0,width=0.75\linewidth]{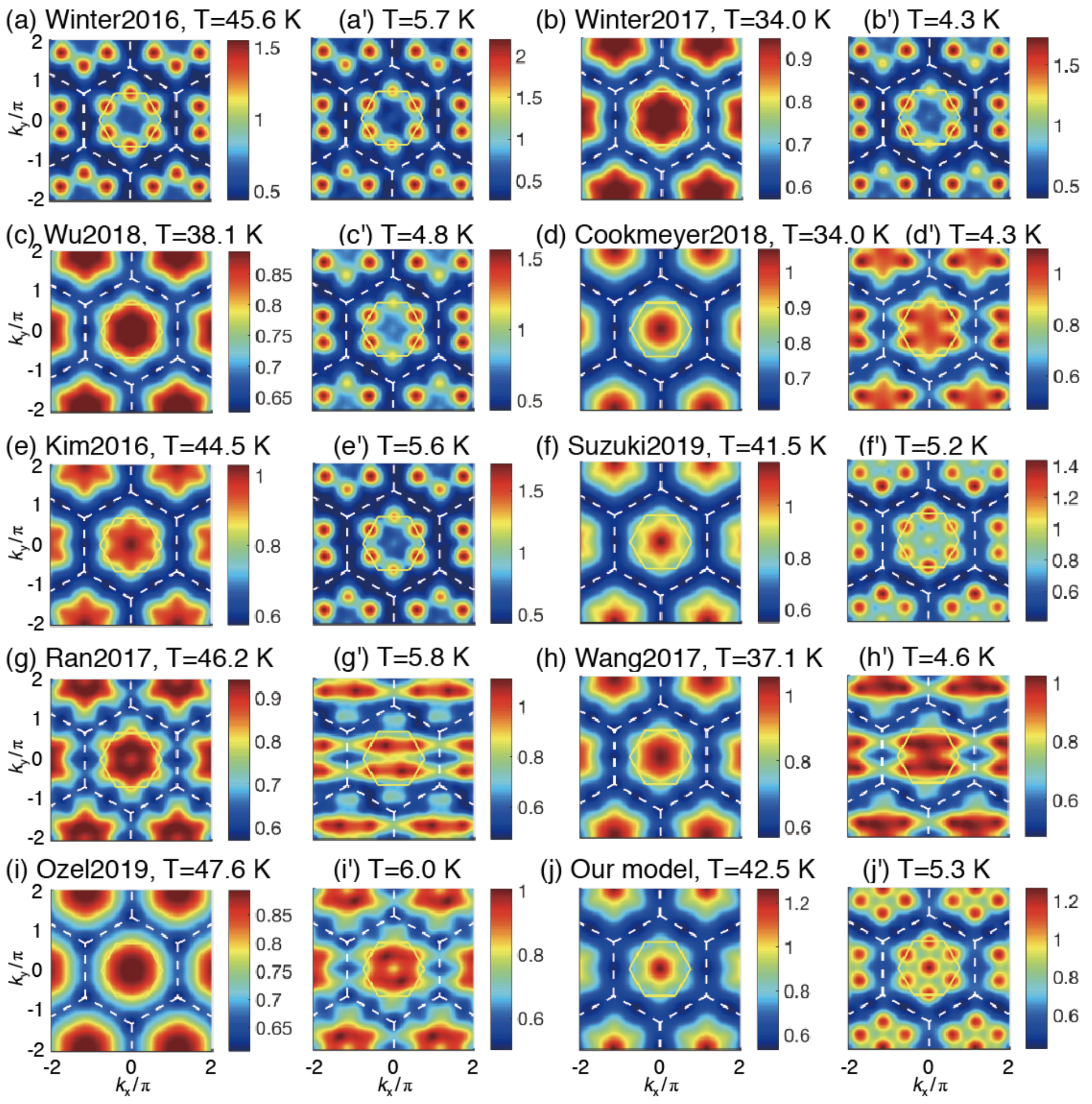}
\renewcommand{\figurename}{\textbf{Supplementary Figure}}
\caption{\textbf{Color maps of the static spin structure factors 
$S(\textbf{k})$ of various candidate \RuCl models.} 
(a-j) show the intermediate-$T$ results
and (a$'$-j$'$) show their low-$T$ counterparts.}
\label{FigS:SSnpj}
\end{figure}

\textbf{Spin structures and the low-$T$ zigzag order.}
The compound \RuCl exhibits a zigzag 
antiferromagnetic order below $T\simeq 7$~K
\cite{Kubota2015,Do2017,Widmann2019,Sears2015,Banerjee2017},
which corresponds to a spin structure peak at the M point. 
Therefore, in Supplementary Fig.~\ref{FigS:SSnpj} we show the static spin structure 
factors $S(\textbf{k}) = \sum_{\gamma = \{x,y,z\}} S^{\gamma \gamma}(\textbf{k})$
at an intermediate $T\sim 40$ K [see Supplementary Fig.~\ref{FigS:SSnpj}(a-j)]
and a low temperature $T\sim 5$ K [Supplementary Fig.~\ref{FigS:SSnpj}(a$'$-j$'$)]. 
Besides the prominent M-peak at low $T$, 
we also anticipate a M-star static structure factor emerging 
at intermediate temperatures, which reflects the short-range 
spin correlations at both the $\Gamma$ and M points of the BZ. 
All these features can be well reproduced 
in the candidate models Winter2017~\cite{Winter2017}, 
Wu2018~\cite{Wu2018}, Cookmeyer2018~\cite{Cookmeyer2018}, 
Kim2016~\cite{Kim2016}, and Suzuki2019~\cite{Suzuki2019}.
As for the dynamical structure factors, it has been discussed before in Ref.~\cite{Laurell2020} 
that only the models Cookmeyer2018~\cite{Cookmeyer2018} and Ran2017~\cite{Ran2017}
could reproduce the M-star shape of the inelastic neutron scattering intensity, 
when integrated over [4.5, 7.5] meV \cite{Banerjee2018}.

Overall, as concluded in Ref.~\cite{Laurell2020}, 
we also find no single candidate model revisited here can satisfactorily 
explain all observed phenomena of experiments on \RuCl. For example, 
both Cookmeyer2018~\cite{Cookmeyer2018} and Ran2017~\cite{Ran2017}
have M-star shape in the intermediate-energy dynamical spin structure,
however, the former does not fit the specific heat and susceptibility 
measurements well and the latter does not host a zigzag oder at low temperature. 
Nevertheless, we note that the model Suzuki2019~\cite{Suzuki2019} can 
reproduce the prominent thermodynamic features such as the double-peaked 
specific heat, highly anisotropic susceptibility, and zigzag static spin structures 
(although the dynamical M star was not found in this candidate model 
according to Ref.~\cite{Laurell2020}). 
Our model, on the other hand, accurately describes the spin interactions 
and explain thus major experimental findings in the compound \RuCl.

\subsection{{\RuCl Model Simulations under In-plane 
\rev{and Tilted} Magnetic Fields}}
\label{SN:IPFields}
In this Note, we show various thermodynamic properties 
of \RuCl under in-plane and tilted $\textbf{c}'$ fields 
(see Fig.~\B{1b} 
in the main text), and compare the 
simulated results to experimental measurements.

\begin{figure}[t!]
\includegraphics[angle=0,width=1\linewidth]{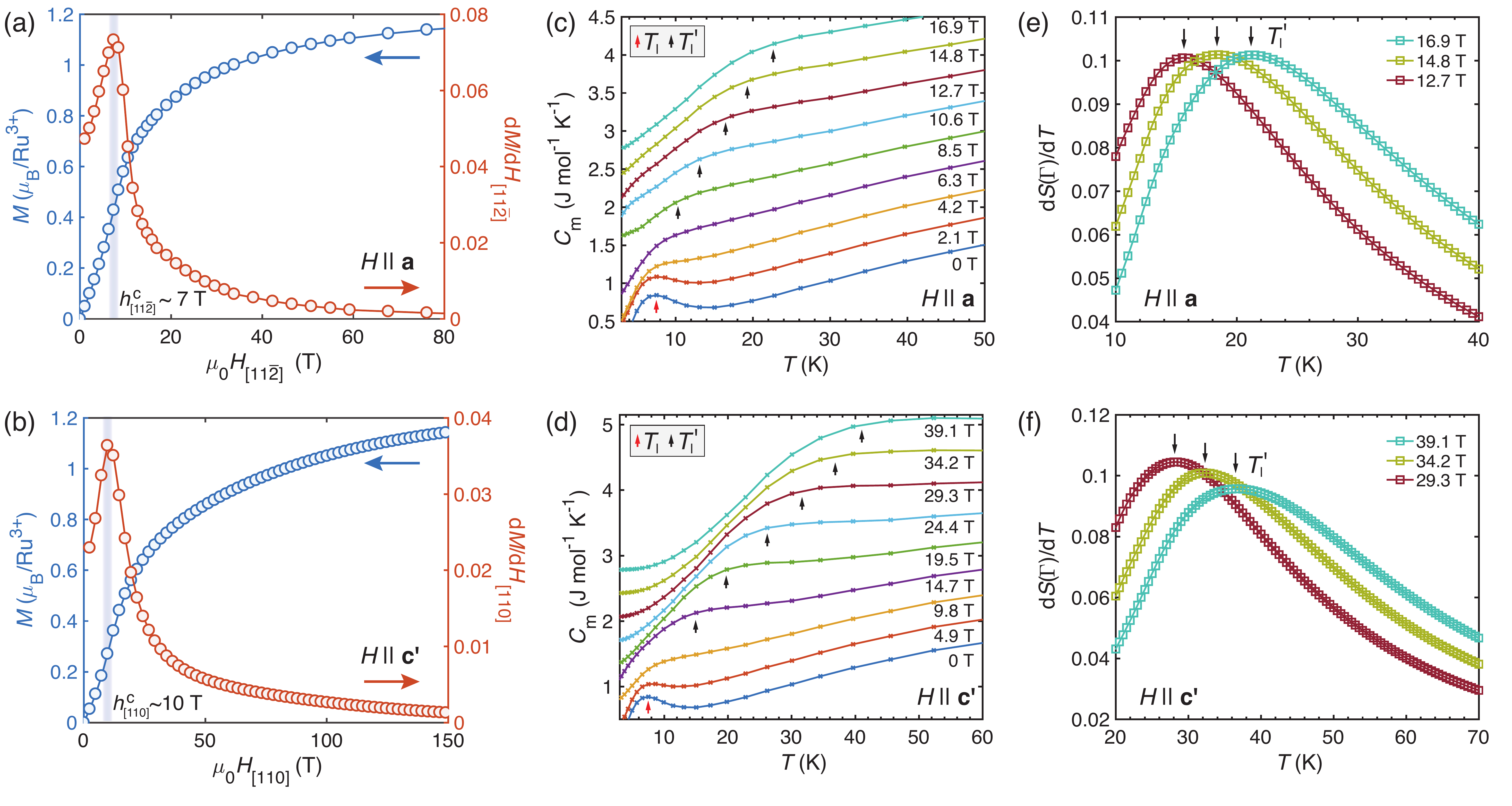}
\renewcommand{\figurename}{\textbf{Supplementary Figure}}
\caption{\textbf{Magnetization curves and thermodynamic 
properties under fields $H_{[11\bar{2}]}\parallel \textbf{a}$ 
and $H_{[110]} \parallel \textbf{c}'$.} The ground-state 
magnetization curves, under (a) in-plane field $H_{[11\bar{2}]}$ 
along $\textbf{a}$- and (b) the tilted $H_{[110]}$ along 
$\textbf{c}'$-axis, are plotted with their derivatives $dM/dH$,
where only one QPT can be observed in both cases.
(c, d) Low-temperature specific heat curves $C_{\rm m}$, 
with each curve shifted by 0.35~J mol$^{-1}$ K$^{-1}$ 
vertically for clarify. The low-$T$ scale $T_{\rm l}$ and $T_{\rm l}'$ 
are indicated by the red and black arrows, respectively.
(e,f) show the temperature derivatives of the uniform 
magnetization, i.e., $dS(\Gamma)/dT$, with the peaks 
located at $T'_{\rm l}$.
}
\label{FigS:H112}
\end{figure}

\textbf{Magnetization curves along the $\textbf{a}$- and $\textbf{c}'$-axis.} 
In Supplementary Fig.~\ref{FigS:H112}(a,b), we show the magnetization 
curves obtained from the $T=0$ DMRG calculations 
on YC4$\times$4$\times$2 geometry. 
QPT takes place 
at 7~T and 10~T, for $H_{[11\bar{2}]}\parallel \textbf{a}$ 
and $H_{[110]} \parallel \textbf{c}'$, respectively.
The QPT is clearly signaled by the divergent 
derivative $d M/d H$, where the zigzag order 
becomes suppressed and the system enters 
the polarized phase. As the field further increases,
the uniform magnetization $M$ gradually 
approaches the saturation value.

\textbf{Specific heat under magetic fields and the
Zeeman energy scale $T'_{\rm l}$.} 
In Supplementary Fig.~\ref{FigS:H112}(c,d), we show the low-$T$ part 
of the magnetic specific heat curves on 
YC4$\times$6$\times$2 systems calculated by XTRG. 
For $\mu_0H_{[11\bar{2}]} < 7$~T, we find the height 
of the $C_{\rm m}$ peaks is suppressed by fields,
and the low-temperature scale $T_{\rm l}$ associated with the
zigzag order, as indicated by the red arrow in Supplementary Fig.~\ref{FigS:H112}(c), 
decreases to zero when the field approaches
the critical value, in agreement with experiments \cite{Kubota2015}.
Above that, a new low-temperature scale $T'_{\rm l}$ builds up
and moves towards higher temperatures almost linearly 
vs $H_{[11\bar{2}]}$, as indicated by the black arrows 
in Supplementary Fig.~\ref{FigS:H112}(c). We find $T'_{\rm l}$ 
is intimately related to the uniform magnetization, as observed
from the temperature variation of spin structure intensity 
$S(\Gamma)$. In Supplementary Fig.~\ref{FigS:H112}(e), 
we show the derivative ${\rm d}S(\Gamma)/{\rm d}T$ 
exhibits clear peaks, whose locations are in agreement 
with the temperature scale $T'_{\rm l}$ determined 
from the low-temperature peak of $C_{\rm m}$ in 
Supplementary Fig.~\ref{FigS:H112}(c). The peaks moving towards higher 
temperature as the magnetic field increases, and we 
thus relate $T'_{\rm l}$ to the Zeeman energy scale.
Beside in-plane fields, as depicted in Supplementary Fig.~\ref{FigS:H112}(d,f),
the case of $H_{[110]} \parallel \textbf{c}'$ exhibits 
very similar behaviors.

\textbf{Energy spectra under in-plane fields.}
In Supplementary Fig.~\ref{FigS:InPSpec} we show ED results of the energy 
spectra on the 24-site cluster (24a), under two in-plane fields 
$H_{[11\bar{2}]}  \parallel \textbf{a}$ and $H_{[1\bar{1}0]} 
\parallel \textbf{b}$ axis [c.f., Supplementary Fig.~\ref{FigS:InPSpec}(a)]. 
The energy spectra results under 
various fields are plotted in Supplementary Fig.~\ref{FigS:InPSpec}(b) and (c), 
for $H \parallel \textbf{a}$ and $\textbf{b}$, respectively.
From there we observe the spin gap $(E_{\rm 1}-E_{\rm 0})$
changes its behavior at an intermediate field, which can be estimated
as the transition field $h_{[11\bar{2}]}^c$ and $h_{[1\bar{1}0]}^c$,
and the spectra appear differently along $\textbf{a}$ and $\textbf{b}$ 
directions, consistent with the observation in recent experiments
\cite{Lampen-Kelley2018,Yokoi2020arXivHalf}.  

\begin{figure}[h!]
\includegraphics[angle=0,width=0.85\linewidth]{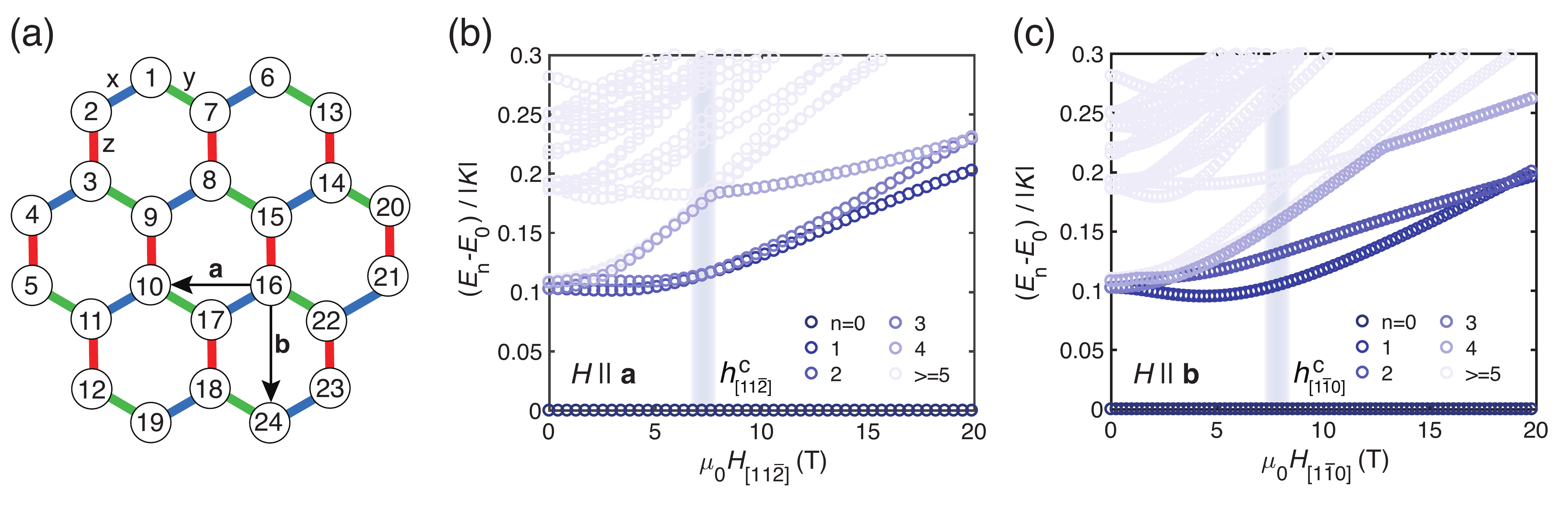}
\renewcommand{\figurename}{\textbf{Supplementary Figure}}
\caption{\textbf{The energy spectra under in-plane magnetic fields.} 
(a) The 24-site cluster involved in the ED calculations, with 
the directions of two in-plane fields $H_{[11\bar{2}]} \parallel \textbf{a}$ 
and $H_{[1\bar{1}0]} \parallel \textbf{b}$ indicated
by the black arrows. (b,c) show the energy spectra 
($E_{\rm n}-E_{\rm 0})$ under the in-plane fields,
with the ground-state energy $E_0$ subtracted.
In two cases, the spin gap starts to increase rapidly 
at about $h_{[11\bar{2}]}^c\simeq7$~T and 
$h_{[1\bar{1}0]}^c \simeq 8$~T, respectively,
as indicated by the thick vertical lines.
}
\label{FigS:InPSpec}
\end{figure}

\begin{figure}[h!]
\includegraphics[angle=0,width=0.8\linewidth]{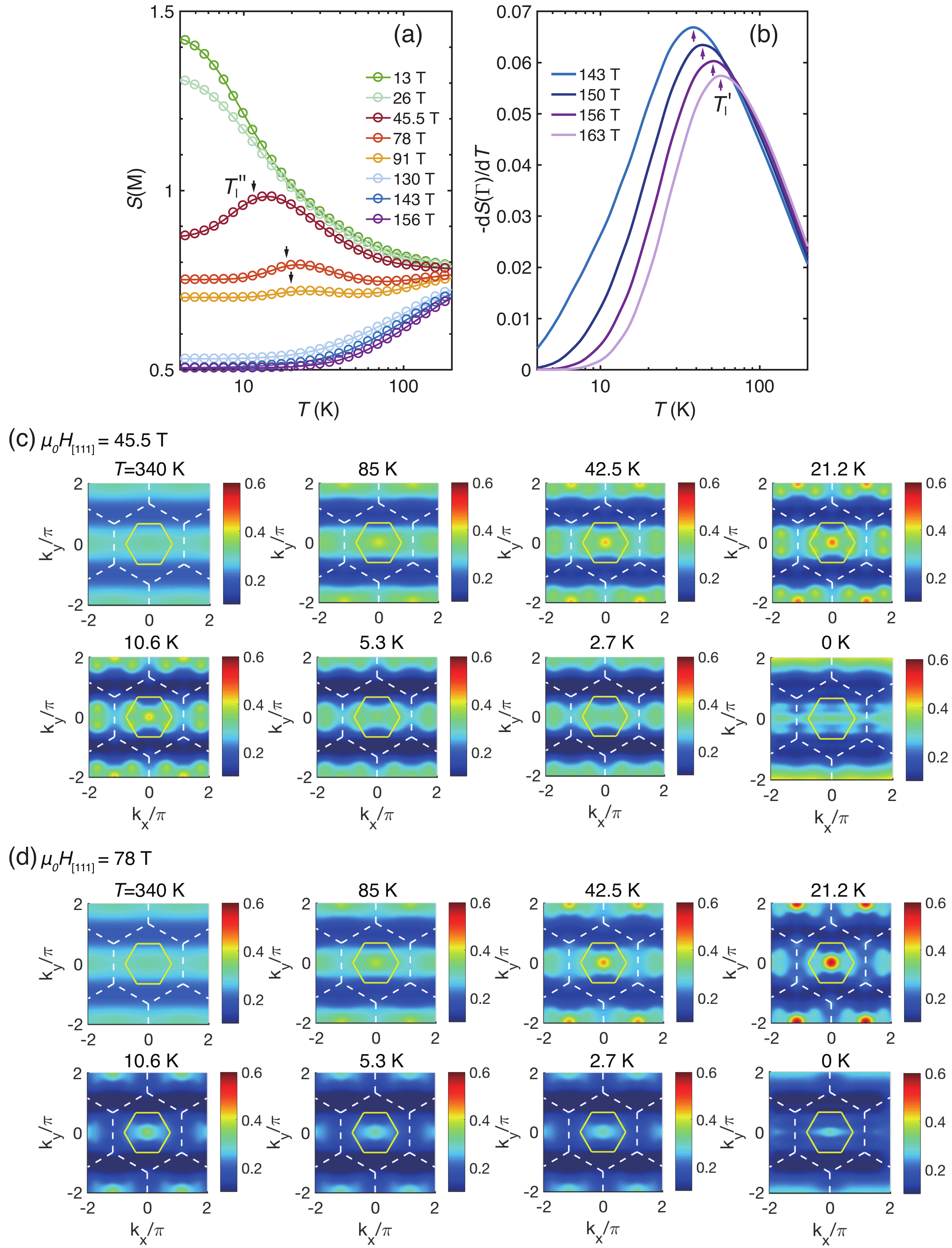}
\renewcommand{\figurename}{\textbf{Supplementary Figure}}
\caption{ \textbf{The spin structure factors under out-of-plane $H_{[111]}$ fields.} 
We show (a) the $S($M$)$ and (b) $-dS(\Gamma)/dT$ curves with 
temperature $T$, which are computed on the YC4$\times$6$\times$2 lattice.
The black arrows in (a) represent the temperature scale $T''_{\rm l}$ 
determined from the low-$T$ peak of the specific heat in Fig.~\B{6} 
of the main text, and the purple ones in (b) indicates the low-$T$ scales $T'_{\rm l}$ 
which locates at the peak of $-dS(\Gamma)/dT$. The static spin-structure 
factors $\tilde{S}^{z z}(\textbf{k})$ [\rev{following the definition in Eq.~(2) 
of the main text}] under the field of $\mu_0H_{[111]}=45.5$~T 
and 78~T are shown in (c) and (d), respectively, from high to low 
temperatures and compared to the ground-state results. 
The data are plotted here in the same colorbar as that in 
Fig.~\B{3e-g} of the main text.
}
\label{FigS:SMG}
\end{figure}

\subsection{{Quantum Spin Liquid under Out-of-plane 
Magnetic Fields}}
\label{SN:QSL}

\textbf{Finite-temperature Spin structures under out-of-plane fields.}
In Supplementary Fig.~\ref{FigS:SMG}, we plot the spin structure factors
at two typical momentum points in the BZ, i.e., 
$\textbf{k}=$M and $\Gamma$. From the 
$S({\rm M})$ curves in Supplementary Fig.~\ref{FigS:SMG}(a),
we find the zigzag order [cf. 13~T and 26~T lines in 
Supplementary Fig.~\ref{FigS:SMG}(a)] gets suppressed 
in the quantum spin liquid (QSL) phase 
under strong out-of-plane magnetic fields
[see, e.g., 45.5~T, 78~T, and 91~T lines in 
Supplementary Fig.~\ref{FigS:SMG}(a)]. 
In the QSL phase, there exists enhanced 
M-point spin correlation near the emergent 
lower temperature scale $T''_{\rm l}$, as indicated
by the black arrows in Supplementary Fig.~\ref{FigS:SMG}(a).
In Supplementary Fig.~\ref{FigS:SMG}(b), we show the temperature
derivative $-dS(\Gamma)/dT$, whose peak position signals 
the low-temperature scale $T'_{\rm l}$, below which the 
field-induced uniform magnetization gets established,
i.e., $T'_{\rm l}$ is a crossover temperature scale to the 
polarized state. As shown in Supplementary Fig.~\ref{FigS:SMG}(b), 
$T'_{\rm l}$ moves to higher temperatures as the field increases. 

In the main text, we have discussed the static spin-structure factors $\tilde{S}^{z z}(\textbf{k})$ 
under an out-of-plane field of 45.5~T (cf. the insets of  Fig.~\B{5b} of the main text). 
In Supplementary Fig.~\ref{FigS:SMG}(c,d), we supplement with XTRG results of $\tilde{S}^{z z}(\textbf{k})$ 
at various temperatures and under two fields, 45.5~T and 78~T. For each case,
we compare the finite-$T$ XTRG results to the $T=0$ data obtained by DMRG.
At high temperature, i.e., $T=340$~K, the spin structure is virtually featureless for both cases, 
with only a faint stripe due to the strong Kitaev interaction ($\sim -25$~meV), 
which becomes more and more clear as the system cools down near the low-$T$
scale $T''_{\rm l}$, around which the brightness at $\Gamma$ and M points are 
slightly enhanced due to the strong spin fluctuations. When $T<T''_{\rm l}$ 
the $\Gamma$ peak in the spin structures becomes flattened, yet the stripy 
background remains distinct, which is in excellent agreement with the DMRG results, 
supporting strongly the existence of QSL state under high out-of-plane fields.

\begin{figure}[!tbp]
\centering
\includegraphics[angle=0,width=0.8\linewidth]{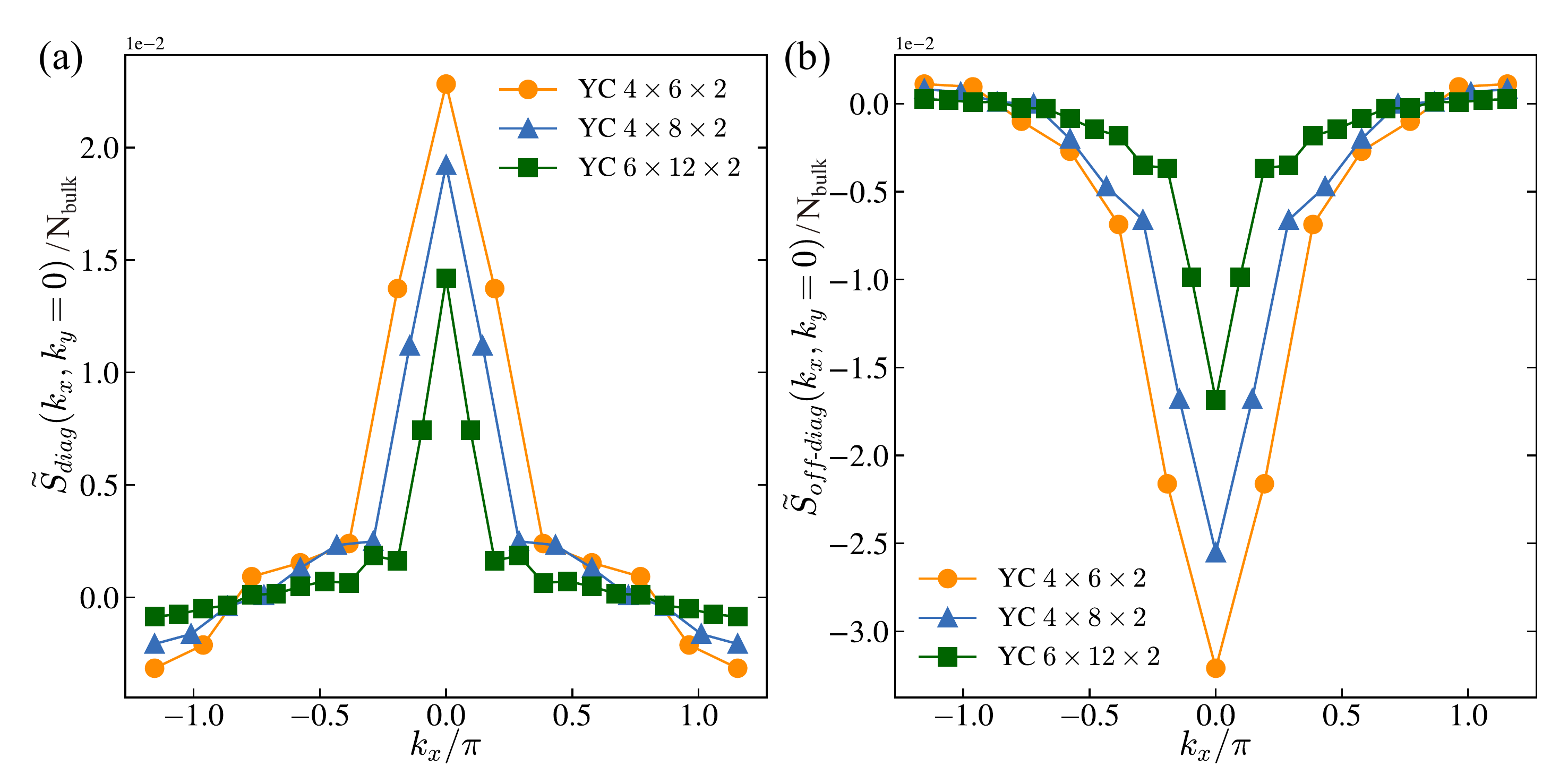}
\renewcommand{\figurename}{\textbf{Supplementary Figure}}
\caption{{\textbf{The static spin structure factors cut at $k_y=0$.} 
(a) shows }$\tilde{S}_{\rm diag}(k_x, k_y=0)/{N_{\rm{bulk}}}$ and (b) 
$\tilde{S}_{\rm off-diag}(k_x, k_y=0)/{N_{\rm{bulk}}}$ in the 
intermediate QSL phase under $\mu_0H_{[111]} 
\simeq 78$~T on cylinders of different sizes.}
\label{FigS:SkCutDMRG}
\end{figure}

\textbf{Ground-state static spin structures under a field of 78~T.}
In Supplementary Fig.~\ref{FigS:SkDMRG} and Supplementary Fig.~\ref{FigS:SkCutDMRG}, 
we show the ground-state static structure factor 
$\tilde{S}^{\alpha\beta}(\textbf{k}) = {\sum_{j\in {\rm bulk},\ j\neq i_0}} 
e^{i\textbf{k}\cdot(\textbf{r}_j-\textbf{r}_{i_0})} (\langle S^{\alpha}_{i_0} 
S^{\beta}_j\rangle -\langle S^{\alpha}_{i_0}\rangle\langle S^{\beta}_{j}\rangle)$
of our \RuCl model on cylinders of different sizes in the high-field QSL phase
(under a magnetic fields of 78~T), obtained by DMRG simulations with 
bond dimension up to $D=2048$. 
{Here we fix a central site $i_0$ and 
compute the correlations with respect to this reference site, instead 
of computing the all-to-all correlations, to speed up the computations.}
From Supplementary Fig.~\ref{FigS:SkDMRG}, we find strong stripy background in 
$\tilde{S}^{xx}$, $\tilde{S}^{yy}$, and $\tilde{S}^{zz}$, 
{divided by 
$N_{\rm{bulk}}$, where $N_{\rm{bulk}}$ is the number of 
bulk sites (i.e., with the left- and right-most columns 
of the cylinder skipped)}, representing intrinsic Kitaev spin liquid
characteristics. There also exists a bright $\Gamma$ point in the BZ, 
and we depict the profiles with fixed $k_y=0$ in Supplementary Fig.~\ref{FigS:SkCutDMRG}, 
to check whether the structure factor diverges at any $\textbf{k}$ point. 
In Supplementary Fig.~\ref{FigS:SkCutDMRG} we find both the diagonal part 
$\tilde S_{{\rm diag}}(\textbf{k})=\sum_{\alpha,\beta}{\tilde 
S^{\alpha\beta}(\textbf{k})\delta_{\alpha\beta}}$ and the off-diagonal 
part $\tilde S_{{\rm off-diag}}(\textbf{k})=\sum_{\alpha,\beta}{\tilde 
S^{\alpha\beta}(\textbf{k})(1-\delta_{\alpha\beta})}$,
{when divided by $N_{\rm bulk}$}, are moving towards zero 
as the system size increases, indicating the absence
of spontaneous long-range order in the phase.

\begin{figure}[!tbp]
\centering
\includegraphics[angle=0,width=0.98\linewidth]{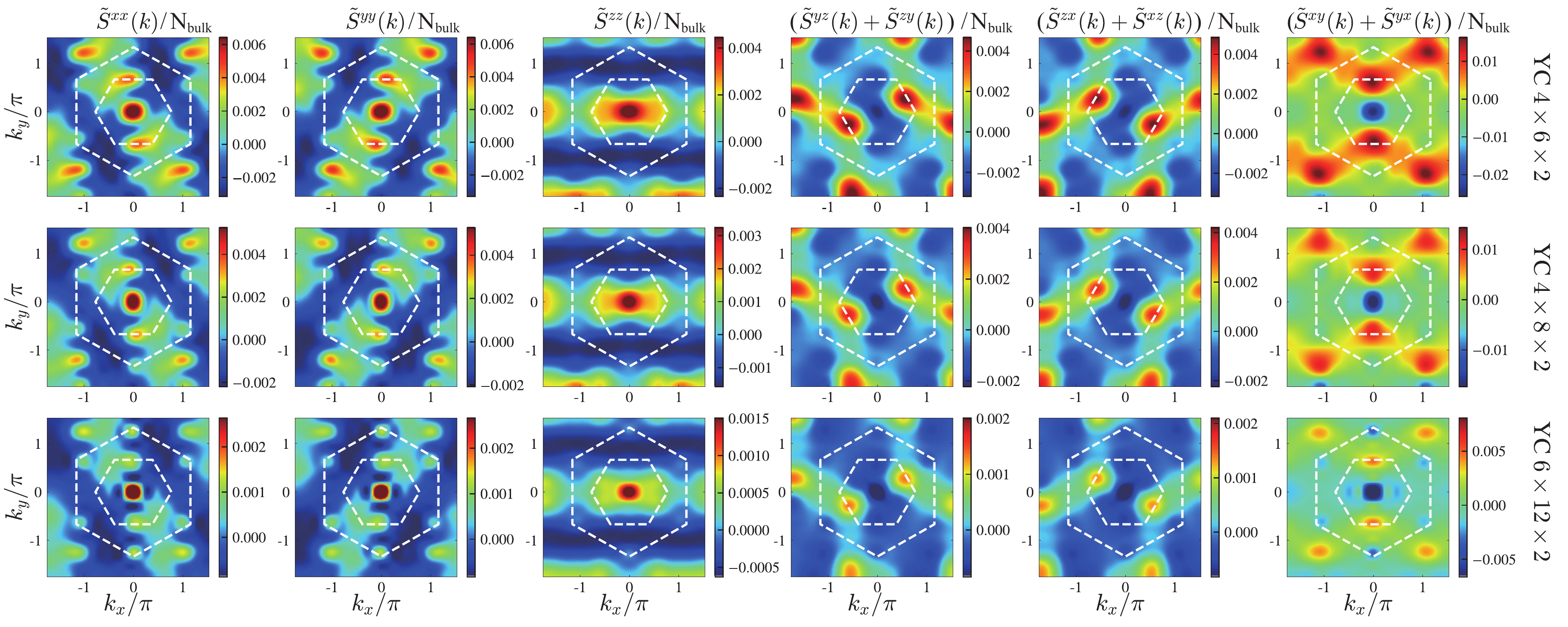}
\renewcommand{\figurename}{\textbf{Supplementary Figure}}
\caption{{\textbf{Color maps of the static spin structure factors $\tilde{S}^{\alpha\beta}(\textbf{k})/{N_{\rm{bulk}}}$.}
The results are calculated} 
in the intermediate QSL phase under $\mu_0H_{[111]}\simeq 78$~T 
on cylinders of different sizes, obtained by DMRG simulations.
}
\label{FigS:SkDMRG}
\end{figure}

\textbf{Ground-state entanglement scaling.}
The nature of the intermediate QSL phase can be further characterized 
by entanglement scaling in the ground state. Here we study the von 
Neumann entanglement entropy $S_v$. Considering a bipartition of 
the system into two parts A and B, the reduced density matrix of A can be 
obtained by tracing out the degrees of freedom of B, i.e., $\rm \rho_A = Tr_B \rho$, 
where $\rho=|\psi_0\rangle\langle \psi_0|$ is density matrix of the normalized ground 
state $|\psi_0\rangle$ of the whole system. 
The von Neumann entropy is defined as 
\begin{equation}
S_v=-\rm{Tr}(\rho_A ln\rho_A).
\end{equation}
Here, we consider the cut along circumference direction and measure $S_v$ 
for each cut at $l$. By virtue of the conformal mapping: 
$l\rightarrow \tilde{l}=(L/\pi)\sin(\pi l/L)$, a central charge $c$ 
on the open cylindrical geometry can be formally extracted using
\begin{equation}
S_v=\frac{c}{6}\ln{\tilde{l}} + {\rm const.},
\label{Eq:CentralCharge}
\end{equation}
originally proposed in the (1+1)-D conformal field theory. The prominent ``dome"-like 
$S_v$ curves in Supplementary Fig.~\ref{FigS:CentralCharge}(a) together with the linear fitting with effective 
central charge $c$ in Supplementary Fig.~\ref{FigS:CentralCharge}(b), indicates a gapless intermediate phase. 
This is consistent with the gapless feature observed in the energy spectra as shown in 
Fig.~\B{5d} and algebraic magnetic specific heat in Fig.~\B{6d} 
of the main text. Furthermore, the effective central charge $c$ 
increases with the system width $W$, i.e. the allowed $\hat{y}$ 
momenta in the Brillouin zone, 
suggesting the increased gapless modes.

\begin{figure}[!tbp]
\centering
\includegraphics[angle=0,width=0.85\linewidth]{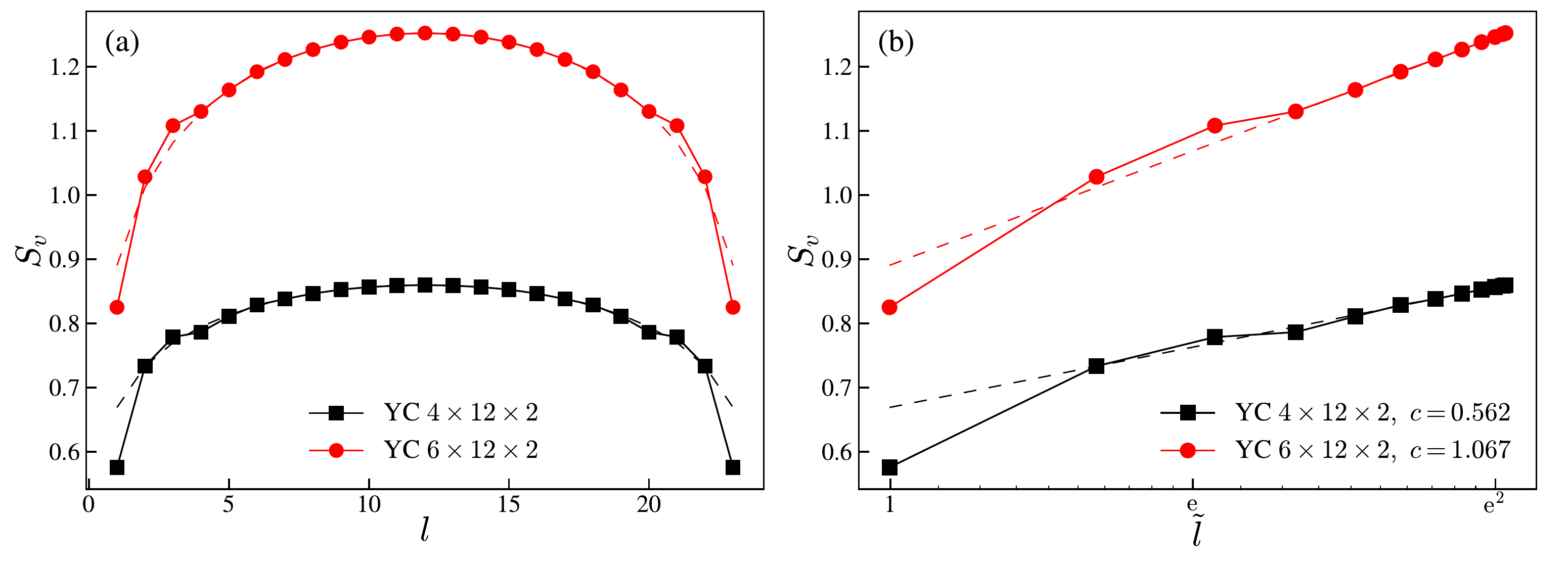}
\renewcommand{\figurename}{\textbf{Supplementary Figure}}
\caption{\textbf{The ground-state entanglement 
    entropy $S_{v}$ in the intermediate phase.} 
    (a) The $S_{v}$ profile over the system with 
    $\mu_0H_{[111]} \simeq 78$~T on cylinders of length $L=12$.
    (b) $S_{v}$ versus conformal distance 
    $\tilde{l}\equiv(L/\pi)\sin(\pi l/L)$ in logarithmic scale. 
    The fittings are plotted as dashed lines in both panels,
    which reveal a central charge $c$ close to 0.5 
    and 1 for $W=4$ and 6, respectively.
}
\label{FigS:CentralCharge}
\end{figure}

\subsection{Variational Monte Carlo Results}\label{SN:VMC}

\textbf{Spin-liquid ansatz based on $D_{\rm 3d}\times Z_2^T$ symmetry.}
We start with the extended Kitaev honeycomb model containing $K$, $J$, $\Gamma$ and $\Gamma'$ interactions. 
The most general expression of the mean-field Hamiltonian ansatz~\cite{Wang2019,GammaPrime,Liu2018} 
with nearest-neighbor couplings reads
\beqn\label{MFPSG}
H_{\rm mf}^{\rm SL} & = & \!\!\!\! \sum_{\langle i,j \rangle \in\alpha\beta(\gamma)} \!\!\! \Big( {\rm Tr}
\, [U_{ji}^{(0)} \psi_i^\dag \psi_j] \! + \! {\rm Tr} \, [U_{ji}^{(1)} \psi_i^\dag
(i R_{\alpha\beta}^\gamma) \psi_j] + {\rm Tr} \, [U_{ji}^{(2)}
\psi_i^\dag \sigma^\gamma \psi_j] \! + \! {\rm Tr} \, [U_{ji}^{(3)} \psi_i^\dag
\sigma^\gamma R_{\alpha\beta}^\gamma \psi_j] \! + \! {\rm H.c.} \Big) \nonumber\\
& & \;\;\;\; +  \sum_i {\rm Tr}  (\pmb \lambda_i\cdot  \psi_i \pmb \tau\psi_i^\dag ),
\eeqn
where $\psi_i = (C_i\ \bar C_i)$, $C_i=(c_{i\dn}, c_{i\up})^T, \bar C_i = (c_{i\downarrow}^\dag, -c_{i\uparrow}^\dag)^T$, 
and $R_{\alpha\beta}^\gamma = - \frac{i}{\sqrt{2}} (\sigma^\alpha + \sigma^\beta)$. Owing to the subtle $SU(2)$ 
gauge symmetry~\cite{Anderson1988} in the fermionic spinon representation, the particle number constraint 
$\hat{N}_i=1$ is extended into the $SU(2)$ gauge invariant three-component form ${\rm Tr}  (\pmb \psi_i \pmb \tau\psi_i^\dag )=0$ 
which are ensured by the three  Lagrangian multipliers $\lambda^{x,y,z}$,
where $\tau^{x,y,z}$ are the generators of the $SU(2)$ gauge group.
The matrices $U_{ji}^{(0,1,2,3)}$ can be expanded with the identity matrix and $\tau^{x,y,z}$, where the expanding coefficients form a subset of $\pmb R$. For the present model, the $SU(2)$ gauge symmetry breaks down to its subgroup $Z_2$ (which is called the invariant gauge group, IGG) and the resultant quantum spin liquids are $Z_2$ QSLs.

A QSL ground state preserves the whole space group symmetry whose point group is $D_{\rm 3d}\times Z_2^T$. However, the symmetry group of a spin liquid mean-field Hamiltonian is the projective symmetry group (PSG)~\cite{igg,You2012} whose group elements are space group operations followed by $SU(2)$ gauge transformations. Although there are more than 100 classes of PSGs for Z$_2$ QSLs, here we adopt the PSG that describes the symmetry of the mean-field Hamiltonian of the pure Kitaev model (called Kitaev PSG). Under this constraint, the allowed mean-field Hamiltonian contains the following parameters.

Firstly, recalling that in KSL the Kitaev interactions are decoupled in a way that the $c$ Majorana fermions are not mixed with the $b^{\gamma}$ Majorana fermions, we have
\begin{eqnarray}\label{Kitaevmf}
H_{\rm mf}^{K} & = & \sum_{\langle i,j \rangle \in\alpha\beta(\gamma)} \rho_a (ic_ic_j) +
\rho_c (i b_i^\gamma b_j^\gamma)  \\
& = & \sum_{\langle i,j \rangle \in\alpha\beta(\gamma)} \Big( i \rho_a {\rm Tr} \big(\psi_i^\dagger \psi_j + \tau^x \psi_i^\dagger \sigma^x \psi_j + \tau^y \psi_i^\dagger \sigma^y \psi_j + \tau^z \psi_i^\dagger \sigma^z \psi_j \big) \notag\\
&&+ i \rho_c {\rm Tr} \big(\psi_i^\dagger \psi_j + \tau^\gamma \psi_i^\dagger \sigma^\gamma \psi_j  - \tau^\alpha \psi_i^\dagger \sigma^\alpha \psi_j - \tau^\beta \psi_i^\dagger \sigma^\beta \psi_j\big) + {\rm H.c.} \Big) \notag
\end{eqnarray}

Similarly, the $\Gamma$ interactions are decoupled as
\beqn\label{Gm}
H_{\rm mf}^{\Gamma} & = & \!\! \sum_{\langle i,j \rangle \in\alpha\beta(\gamma)} \!\! i\rho_d
(b_i^\alpha b_j^\beta + b_i^\beta b_j^\alpha) \\ & = & \!\! \sum_{\langle i,j \rangle \in\alpha\beta(\gamma)} \!\! \Big( i\rho_d {\rm Tr} \big( \tau^\alpha \psi_i^\dag
\sigma^\beta \psi_j + \tau^\beta \psi_i^\dag \sigma^\alpha \psi_j\big)
 + {\rm H.c.} \Big), \notag
\eeqn
and the $\Gamma'$ interactions are decoupled as
\beqn\label{Gmp}
H_{\rm mf}^{\Gamma'} & = & \!\! \sum_{\langle i,j \rangle \in\alpha\beta(\gamma)} \!\! i\rho_f
(b_i^\alpha b_j^\gamma + b_i^\gamma b_j^\alpha + b_i^\beta b_j^\gamma + b_i^\gamma b_j^\beta) \\
& = & \!\! \sum_{\langle i,j \rangle \in\alpha\beta(\gamma)} \!\! \Big( i\rho_f {\rm Tr} \big( \tau^\alpha \psi_i^\dag
\sigma^\gamma \psi_j + \tau^\gamma \psi_i^\dag \sigma^\alpha \psi_j + \tau^\beta \psi_i^\dag \sigma^\gamma \psi_j 
+ \tau^\gamma \psi_i^\dag \sigma^\beta \psi_j \big) +  {\rm H.c.} \Big). \nonumber
\eeqn

Comparing with the general form Supplementary Eq.~(\ref{MFPSG}),  the decouplings expressed in
Supplementary Eqs.~(\ref{Kitaevmf}), (\ref{Gm}) and (\ref{Gmp}) contribute the terms
$ {\tilde U}_{ji}^{(0)} = i (\rho_a + \rho_c), \ {\tilde U}_{ji}^{(1)} = i (\rho_a - \rho_c  + \rho_d + 2\rho_f) (\tau^\alpha + \tau^\beta),\  {\tilde U}_{ji}^{(2)} = i (\rho_a + \rho_c) \tau^\gamma + i \rho_f(\tau^\alpha +\tau^\beta ), \  {\tilde U}_{ji}^{(3)} = i (\rho_c - \rho_a - \rho_d) (\tau^\alpha - \tau^\beta ),$
to the coefficients $U_{ji}^{(m)}$, in which $j$ and $i$ specify $\gamma$.

Secondly, we consider a more general mean-field decoupling which means $c$ Majorana fermions can mix with $b^{\gamma}$ Majorana fermions. In addition, the most general coefficients preserving the $C_3$ rotation symmetry (in the PSG sense) also contain multiples of the uniform ($I$) and $\tau^x + \tau^y + \tau^z$ gauge components,
$ {\tilde {\tilde U}_{ji}^{(0)}} = i \eta_0 + \eta_1 (\tau^x + \tau^y + \tau^z), \ {\tilde {\tilde U}_{ji}^{(1)}} = \eta_2 + i \eta_3 (\tau^x + \tau^y + \tau^z), \ {\tilde {\tilde U}_{ji}^{(2)}} = \eta_4 + i \eta_5 (\tau^x + \tau^y + \tau^z), \  {\tilde {\tilde U}_{ji}^{(3)}} = \eta_6 + i \eta_7 (\tau^x + \tau^y + \tau^z).$
If the full symmetry group, $G = D_{\rm 3d} \times Z_2^T$, is preserved, then only three parameters
$\eta_0$, $\eta_3$, and $\eta_5$ are allowed. Thus a spin-liquid ansatz that preserves the full Kitaev PSG contains the variables $U_{ji}^{(m)} = {\tilde U}_{ji}^{(m)} + {\tilde {\tilde U}_{ji}^{(m)}}$, namely
\begin{equation}\label{Eta}
\begin{aligned}
&U_{ji}^{(0)} = i (\rho_a + \rho_c) +  i \eta_0, \\
&U_{ji}^{(1)} = i (\rho_a - \rho_c  + \rho_d + 2\rho_f) (\tau^\alpha + \tau^\beta)+ i \eta_3 (\tau^x + \tau^y + \tau^z),\\
&U_{ji}^{(2)} = i (\rho_a + \rho_c) \tau^\gamma + i \rho_f(\tau^\alpha +\tau^\beta )+ i \eta_5 (\tau^x + \tau^y + \tau^z), \\
&U_{ji}^{(3)} = i (\rho_c - \rho_a - \rho_d) (\tau^\alpha - \tau^\beta ),
\end{aligned}
\end{equation}
with seven real parameters, $\rho_a$, $\rho_c$, $\rho_d$, $\rho_f$, $\eta_0$, $\eta_3$ and $\eta_5$.

\textbf{Magnetically ordered states.}
Because fermions do not condense alone to form a magnetic order,
we introduce the classical order under single-${\pmb Q}$ approximation~\cite{singleQ} 
to describe the magnetic order of the spin-symmetry-breaking phases 
of the $K$-$J$-$\Gamma$-$\Gamma'$ model.
\[\pmb{M}_i = M \{\sin \phi [\hat {\pmb e}_x \cos (\pmb{Q} \cdot
\pmb{r}_i) + \hat{\pmb e}_y \sin(\pmb{Q} \cdot \pmb{r}_i)] + \cos \phi
\, \hat{\pmb e}_z \},\]
where  $\pmb Q$ is the ordering momentum, $\hat {\pmb e}_{x,y,z}$ 
are the local spin axes (not to be confused with the global spin axes),
and $\phi$ is the canting angle. $\pi/2-\phi$ describes the angle by 
which the spins deviate from the plane spanned by $\hat {\pmb e}_x$ and $\hat {\pmb e}_y$. 
The classical ground state is obtained by minimizing the energy of the trial states.

In our VMC calculations, the static order is treated as a background 
field coupling to the spins as site-dependent Zeeman field,
hence the complete mean-field Hamiltonian for the $K$-$J$-$\Gamma$-$\Gamma'$ model reads
\begin{equation}\label{Order}
H_{\rm mf}^{\rm total} = H_{\rm mf}^{\rm SL} - {\textstyle \frac{1}{2}} \sum_i
(\pmb {M}_i \cdot C_i^\dagger \pmb \sigma C_i + {\rm H.c.})
\end{equation}

The ordering momentum $\pmb Q$ of $\pmb M_i$
in VMC is adopted from the classical
ground state or the classical metastable states
(depending on the energy of the projected state).
For a given $\pmb Q$, the local axes
$\hat {\pmb e}_{x,y,z}$ are fixed as
they are in the classical state, $M$ and $\phi$
are treated as variational parameters.
The amplitude of the static magnetic order $M$
we compute includes quantum corrections.

Finally, since the magnetic order essentially breaks the $C_3$ symmetry, 
the parameter $\eta_7$ and an additional parameter $\theta$ describing 
the ratio of certain parameters in the $z$-bond and $x$-($y$-) bonds are allowed 
(similar to the model discussed in Ref.~\cite{Aniso}).

\textbf{Field-induced chiral spin liquid.} By comparing various ansatzes, including the spin-liquid and magnetically ordered states, our VMC calculations reveals an exotic phase under the out-of-plane magnetic field $\pmb B \equiv \mu_0 H_{[111]}$. As the zigzag order being suppressed by a large field, an intermediate chiral spin liquid (CSL) phase with Chern number $\nu=2$ emerges before the system enters the polarized phase. The phase transition between the zigzag ordered state and the the CSL phase is of first order. According to Kitaev's seminal work~\cite{Kitaev2006}, the CSL is an Abelian phase whose quasiparticle excitations include $a,\bar a,\varepsilon,1$, where 1 denotes the vacuum, $\varepsilon$ is the fermion, $a$ and $\bar a$ are two types of vortices with topological spin $e^{i \pi/4}$. The fusion rules are $\varepsilon\times\varepsilon=1$, $a\times \bar a=1, a\times \varepsilon=\bar a, \bar a\times \varepsilon=a, \bar a\times \bar a=\varepsilon$. The edge of the CSL state is gapless and contains 2 branches of chiral Majorana excitations, each branch carries a chiral central charge of $1/2$. Therefore, the thermal Hall conductance is integer quantized, with $\kappa_{xy}/T = \pi k_B^2/ 6h$ at low temperature.

In the following, we discuss whether the field-induced Z$_2$ CSL is topologically nontrivial or not.
The confinement or deconfinement of the Z$_2$ gauge field is reflected
in the ground state degeneracy (GSD) of the Gutzwiller projected state
when placed on a torus. Therefore, we calculate the density matrix of 
the projected states
from the wave-function overlap $\rho_{\alpha\beta} =
\langle P_G \psi_\alpha | P_G \psi_\beta \rangle = \rho_{\beta\alpha}^*$,
where $\alpha,\beta \in \{++,+-,-+,--\}$ are the boundary conditions
($+$ stands for the periodic boundary condition and $-$
for antiperiodic boundary condition) along $\pmb a_1$
and $\pmb a_2$ direction, respectively.
If $\rho$ has only one significant eigenvalue, with the others vanishingly small,
then the GSD is 1, indicating that the Z$_2$ gauge field is confined.
Otherwise, for $\rho$ with more than one (near-degenerate) nonzero eigenvalues,
then the GSD is nontrivial and hence the Z$_2$ gauge fluctuations are deconfined.
In the VMC calculations, we find the field-induced CSL (with Chern number $\nu=2$) 
is deconfined,  with GSD equals 4
(the eigenvalues of the overlap matrix $\rho$ are given by $0.5915,0.8066,1.0419,1.5600$
under $g_{c^*}\mu_{\rm B} B/|K|=0.52$ for a system on an 8$\times$8$\times$2 torus).
Actually, we have performed a finite-size scaling calculation (not shown),
which indeed indicates that the GSD of this state is 4 in the large-size limit.
Indeed, such a GSD matches the number of topologically distinct quasiparticle types.

\end{document}